\documentclass[letterpaper,twocolumn,10pt]{article}
\PassOptionsToPackage{table}{xcolor}
\usepackage[hyphens]{url}
\usepackage{usenix-2020-09}

\usepackage[utf8]{inputenc}
\usepackage[T1]{fontenc}
\usepackage{amsmath}
\usepackage{amsfonts}
\usepackage{amssymb}

\usepackage{graphicx}
\usepackage{xspace}

\usepackage{hyperref}
\usepackage{caption}
\usepackage{subcaption}

\usepackage{booktabs}
\usepackage{tabularx}
\usepackage{multirow}
\usepackage{array}
\usepackage[normalem]{ulem}

\graphicspath{{./figures/}}

\renewcommand{\paragraph}[1]{\vspace{0.5em}\noindent\textbf{#1.}}

\newcommand\new[1]{#1}

\usepackage{pifont}
\let\oldding\ding%
\renewcommand{\ding}[2][1]{\scalebox{#1}{\oldding{#2}}}
\newcommand{\one}{\ding[1.2]{192}\xspace}
\newcommand{\two}{\ding[1.2]{193}\xspace}
\newcommand{\three}{\ding[1.2]{194}\xspace}
\newcommand{\four}{\ding[1.2]{195}\xspace}

\newcolumntype{L}{>{\raggedright\let\newline\\\arraybackslash\hspace{0pt}}X}

\newcommand{\numMale}{823}

\newcommand{\numTotalParticipants}{2,673}
\newcommand{\numFemale}{715}
\newcommand{\numOther}{35}

\newcommand{\numExcluded}{6}

\title{URL Inspection Tasks: Helping Users Detect Phishing Links in Emails}

\usepackage{paralist}
\author{}

\begin{document}

	\author{
	{\rm Daniele Lain}\\
	ETH Zurich
	\and
	{\rm Yoshimichi Nakatsuka}\\
	ETH Zurich
	\and
	{\rm Kari Kostiainen}\\
	ETH Zurich
	\and
	{\rm Gene Tsudik}\\
	UC Irvine 
	\and
	{\rm Srdjan Capkun}\\
	ETH Zurich
	} 

	\maketitle
	
	\begin{abstract}
		
The most widespread type of phishing attack involves email messages with links pointing to malicious content.
Despite user training and the use of detection techniques, these attacks are still highly effective. 
Recent studies show that it is user {\em inattentiveness}, rather than lack of education, that is one of the key factors in successful phishing attacks. 
To this end, we develop a novel phishing defense mechanism based on \textit{URL inspection tasks}: small challenges (loosely inspired by CAPTCHAs) that, to be solved, require users to interact with, and understand, the basic URL structure. 
We implemented and evaluated three tasks that act as ``barriers'' to visiting the website: (1) correct click-selection from a list of URLs, (2) mouse-based highlighting of the domain-name URL component, and (3) re-typing the domain-name. 
These tasks follow best practices in security interfaces and warning design. 

We assessed the efficacy of these tasks through an extensive on-line user study with \numTotalParticipants{} participants from three different cultures, native languages, and alphabets. 
Results show that these tasks significantly decrease the rate of successful phishing attempts, compared to the baseline case. 
Results also showed the highest efficacy for difficult URLs, such as typo-squats, with which participants struggled the most.
This highlights the importance of (1) slowing down users while focusing their attention and (2) helping them understand the URL structure (especially, the domain-name component thereof) and matching it to their intent.

	\end{abstract}
	
	\section{Introduction}
	\label{sec:introduction}
	Phishing is a widespread problem, with attackers using increasingly sophisticated techniques to deceive users~\cite{lin2022phish}  
-- a situation only exacerbated by the COVID-19 pandemic with its shift to remote work and digital communication~\cite{al2022covid}, 
as well as by increasingly accessible and sophisticated AI tools~\cite{mink2022deepphish} that can
generate highly realistic, yet deceptive, content.

A common goal of phishing attacks is to deliver a \textit{payload} to its victim, usually malicious attachments or URLs pointing
to malicious content (e.g., websites that harvest credentials or install malware) via email~\cite{cofense2023}. 
In the case of malicious attachments, many technical and user-interface countermeasures have been widely 
studied~\cite{aslan2020comprehensive,bravo2010bridging} and deployed. This is further aided by better user education: 
informed users learn to avoid opening attachments from unknown sources~\cite{vishwanath2018suspicion}.

However, for URLs, the situation is different: technical countermeasures (e.g., URL blacklists, machine 
learning techniques) lag behind attackers' increasing sophistication~\cite{oest2020phishtime,sheng2009empirical}.
Mainstream user interfaces offer little help to users besides showing the full URL in address bars and small tool-tips.
Unsurprisingly, URLs are currently the main vector for phishing attacks, more so than attachments~\cite{cofense2023}, 
for the purposes of harvesting credentials and malware delivery~\cite{cofense2023}. It seems that users struggle with 
the current state-of-the-art anti-phishing methods which fail to support their decision-making in: (i) paying attention 
to the URL which they click; (ii) understanding its structure (e.g., what is the domain-name component and what it means); 
and (iii) deciding whether the clicked URL will take them to the website they expect. Recent studies show that 
user inattention is among the main contributors to the success of phishing attacks~\cite{greene2018user,vishwanath2011people}.

Motivated by this, we design and evaluate several \textit{URL inspection tasks}: small challenges 
served to users (when they click on links contained in emails) that must be solved correctly
before they can continue to their (intended) destination.
Solving these challenges requires interaction with the URL: they focus users' attention on the URL they are about to 
visit and require a basic understanding of its structure to be solved. They also help users check if the URL they are 
about to visit matches their intent by (indirectly) making them solve the challenge incorrectly in case of a misunderstanding.
Since solving these challenges requires users to determine where {\em they think they are going}, those who are confused 
by the common impersonation tactics of phishing URLs (e.g., containing the name of a reputable domain to inspire trustworthiness) 
answer incorrectly, thus triggering a warning.

We implemented three types of inspection tasks using three basic HCI mechanisms: \textit{click-selection} among a list of 
candidate URLs, 
\textit{highlighting} the domain by selecting it, and \textit{re-typing} the domain that the user thinks they want to visit.
We evaluated these tasks in a large (\numTotalParticipants{} participants) on-line study, designed as a realistic role-play 
experiment wherein participants pretend to be employees of a fictitious company who routinely manage email in a custom mailbox.
Our results show that inspection tasks prevented participants from falling for phishing attacks, 
compared to a control group that reflected the typical current experience of Internet users, as the rate of successful 
phishing emails fell from 74\% to 35\%.

The studied tasks also outperformed a passive baseline (57\% phishing success rate) where the URL is shown again though
the user only has to confirm their intention to proceed. This testifies to the effectiveness and importance of 
active engagement with the task and its prevention of habituation. 
They also outperformed a \textit{semi-active} baseline (61\% phishing success rate) where participants drag-and-drop parts of the 
URL back in place (thus presenting an active task component) and can only be solved correctly. This approach does not help users 
understand whether the destination matches their intent, while our results demonstrate the importance of this last step.
The difficulty of detecting different types of phishing URLs varies: while our tasks outperformed the baselines for all types 
of URLs that impersonate the victim's domain, they were especially effective against typo-squat URLs 
(that participants struggled with otherwise), decreasing the successful phishing rate from 79\% to 17\%.

\new{Active approaches such as ours should be used sporadically, similar to how CAPTCHAs are used today, and are recommended in scenarios requiring higher security, such as corporate environments.
Indeed, this approach trades off increased vigilance and detection against a moderate increase in user burden and false positives. Our study also aims to understand this tradeoff: our tasks slowed users by 7-10 seconds on average and somewhat increased annoyance compared to a regular email workflow, but provided a higher level of protection.}

The contributions of this work are:
\begin{compactitem}
    \item The concept of \textit{inspection tasks} upon clicking on links in emails to help focus users' attention, 
    and verify whether the URL they are about to visit matches their intent.
    \item Assessment of three types of inspection tasks grounded in basic HCI mechanisms: clicking, highlighting, and re-typing. 
    We tested them on a wide range of phishing URLs as part of a large on-line study with \numTotalParticipants{} participants 
    from the United States, Germany, and Japan.
    \item Results of the study show significant improvement in lowering the fraction of users who succumb to phishing attacks.
    The tasks are especially effective against sophisticated typo-squatting URLs. This effectiveness is due to both (1) 
    active user engagement with the task and (2) helping users check whether a given URL matches their intent.
\end{compactitem}

	\section{Active Tasks for URL Inspection}
	\label{sec:motivation}
	\paragraph{Motivation}
Phishing by email generally uses two attack vectors: URLs and attachments.
One major reason why URL-based email phishing succeeds is due to the difficulty of parsing the complicated 
structure of embedded links by users~\cite{albakry2020url}, and inattentiveness. This is especially the case when malicious 
URLs impersonate legitimate services~\cite{reynolds2020measuring}.
Attachment-based phishing is currently less effective since many modern email clients, browsers, and OSes 
implement defense mechanisms, e.g., via blocking downloads or explicit warnings. Also, 
users have become increasingly aware of the perils of opening unknown attachments~\cite{vishwanath2018suspicion}.

It is thus surprising that for URLs, users are left on their own: both standalone and browser-based 
email clients do not provide much help besides showing the destination of links on small tooltips upon mouse hover. Browsers help users by highlighting URLs in the address bar or hiding their path. However, these countermeasures do not seem to assuage users' 
struggles~\cite{lin2011does,xiong2017domain,dhamija2006phishing}.
Furthermore, users who already made up their minds about a given email~\cite{lain2021phishing} 
might ignore the URL when it is displayed in the browser~\cite{lin2011does}.

Therefore, research has focused on improving email clients and browsers by helping users 
understand links and URLs. This has been done by providing tooltips with information about the 
URLs~\cite{althobaiti2021don}, introducing delays before opening the link~\cite{volkamer2017user}, 
or forcing users to click on it again~\cite{petelka2019put}. Another popular countermeasure employed by 
many online services is a warning page displayed upon clicking on a URL and asking the user to 
confirm that they wish to visit it. 

\begin{figure*}[t]
    \centering
    \includegraphics[width=.8\textwidth]{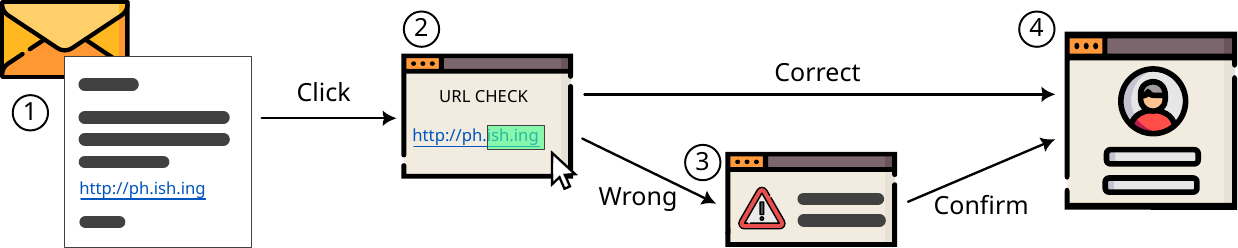}
    \caption{\textbf{Overview of active tasks for URL inspection.} Upon clicking on a link in an 
    email, the user is presented with a task to be solved on the clicked URL, forcing attention and helping to understand where it is taking them.}
    \label{fig:overview}
\end{figure*}

\paragraph{Limitations of Prior Approaches}
We focus on a concrete class of phishing attempts: consider an email containing at least one URL that 
impersonates (resembles) a legitimate website such as
\texttt{example.com.scam.com}, hosted on the attacker-owned \texttt{scam.com} domain, and attempting to impersonate the legitimate website \texttt{example.com}
This is a popular form of URL impersonation~\cite{reynolds2020measuring,zeng2021winding} that aims to deceive 
the user into thinking the bogus URL leads to a legitimate website.
To make an informed decision about the legitimacy of this URL, several things need to happen:

First, a user must take the time to pay attention instead of simply clicking it, which means visually parsing
the URL. An artificial slow-down of user interaction~\cite{volkamer2017user,petelka2019put} here makes sense, since phishing susceptibility is 
often based on quick decisions~\cite{purkait2014empirical}. 

Second, a user needs to {\bf understand} that the URL leads to \texttt{scam.com}. However, just providing additional 
contextual information~\cite{volkamer2017user,nicholson2017can} can be easily ignored~\cite{neupane2015multi}.
Meanwhile, an additional cognitive effort imposed on users makes them less vulnerable to 
phishing~\cite{wang2012research,purkait2014empirical}. 

Third, a user needs help understanding that clicking will NOT lead to the expected website \texttt{example.com}.

\new{Finally, the user needs to know that \texttt{example.com} is the correct domain of their desired service ``Example'', and that they wanted to visit this website instead.}

\subsection{Overview}
Ideally, an effective anti-phishing technique must employ best practices of security interface design: 
(1) prevent habituation and desensitization~\cite{krol2012don}, and (2) require user 
interaction~\cite{bravo2013your} while (3) providing actionable information to help users
make a decision~\cite{li2007usability,schaub2015design,bauer2013warning}. Prior techniques
do not satisfy this.

Our work uses all three aforementioned elements. It involves active challenges that need to be solved 
by interacting with the URL, thus alerting users and directing their attention. It
requires basic understanding of the URL structure to make users understand where a URL would
lead if they were to click it, thus helping users in making a informed decision.
Furthermore, challenges can be designed to (indirectly) help users answer the question \textit{``Where would this URL take you?''}
and notify them in case there is a mismatch between their stated intention and the URL. 
In our example above, a challenge would result in the answer: \texttt{example.com}, 
and warn the user that the URL would in fact bring them to \texttt{scam.com}.

We overview our approach in Figure~\ref{fig:overview}: upon clicking on a link in an email 
(denoted as \one), the user is immediately presented with an attention-enhancing task (denoted as 
\two) that motivates them to inspect and understand the URL, e.g., in a tooltip or a page 
on their browser. There, the URL is presented in a way that is easy to read and understand~\cite{franz2021sok}.
The user has to solve the task correctly in order to proceed to website \four. If 
they make a mistake, an error is shown (\three) by, e.g., presenting both 
the original domain and the user's answer. The user is then asked to confirm whether 
they want to proceed.

\new{Note that our approach alone does not help with user knowledge of the domain for any expected service: we discuss the implications of this gap further in Section~\ref{sec:discussion.approach}.}

	\section{Tasks Design}
	\label{sec:design}
	We face several challenges in creating concrete actionable tasks.
First, we need to understand which tasks can help users and how, plus analyze inherent 
trade-offs between ease-of-use, solving speed, and effectiveness, similar to challenges faced by 
CAPTCHA mechanisms~\cite{searles2023empirical}. Second, it is unclear how to design tasks that 
help users understand URLs, especially phishing URLs, as well as how to understand user intent 
and trigger an error in case of a mismatch. To tackle these challenges, we begin by exploring  
the ecosystem of phishing URLs and then propose a set of appropriate tasks.

\subsection{Types of Phishing URLs}
\label{sec:design.url}
It is important to identify common phishing URL types, since tasks should help users understand 
the URL structure and (hopefully) capture their intentions by triggering an error in case of a 
misunderstanding. Following taxonomies from the 
literature~\cite{reynolds2020measuring,zeng2021winding,tupsamudre2019everything,aung2019survey,canova2015nophish}, 
we observe that there are two main families of phishing URLs: (i) URLs that have no relationship with what they 
are impersonating, e.g., the domain name refers to a compromised domain, a random name, or an IP address; and 
(ii) URLs that somehow refer to what they are impersonating. Type-(i) can be spotted by a user by simply re-reading 
the URL to realize that it does not correspond to their intent. However, type-(ii) is more deceiving since the URL
contains a literal, near-literal (e.g., a typosquat), or partial name of the domain is impersonates, making it more  
difficult to understand~\cite{albakry2020url}. We thus focus on type-(ii).

The impersonated domain can appear in different parts of a phishing URL~\cite{reynolds2020measuring,zeng2021winding,canova2015nophish}.
Suppose that an adversary wants to impersonate \textit{example.com}. The impersonation can occur in the 
following parts (actual domain is underlined):
\begin{compactitem}
    \item \textbf{Subdomains:} \texttt{example.\uline{com-login.com}}.
    \item \textbf{Beginning of Domain:} \texttt{\uline{example-login.com}}.
    \item \textbf{End of Domain:} \texttt{\uline{login-example.com}}.
    \item \textbf{In Path:} \texttt{\uline{login.com}/example.com}.
    \item \textbf{Typosquat:} \texttt{\uline{exampie.com}} that substitutes the character \texttt{l} with the similar-looking \texttt{i}.
\end{compactitem}
In the next section, we discuss the potential impact of each task over the different URL types.

\begin{figure*}[!t]
    \centering
    \begin{subfigure}{0.33\textwidth}
        \centering
        \includegraphics[width=\textwidth]{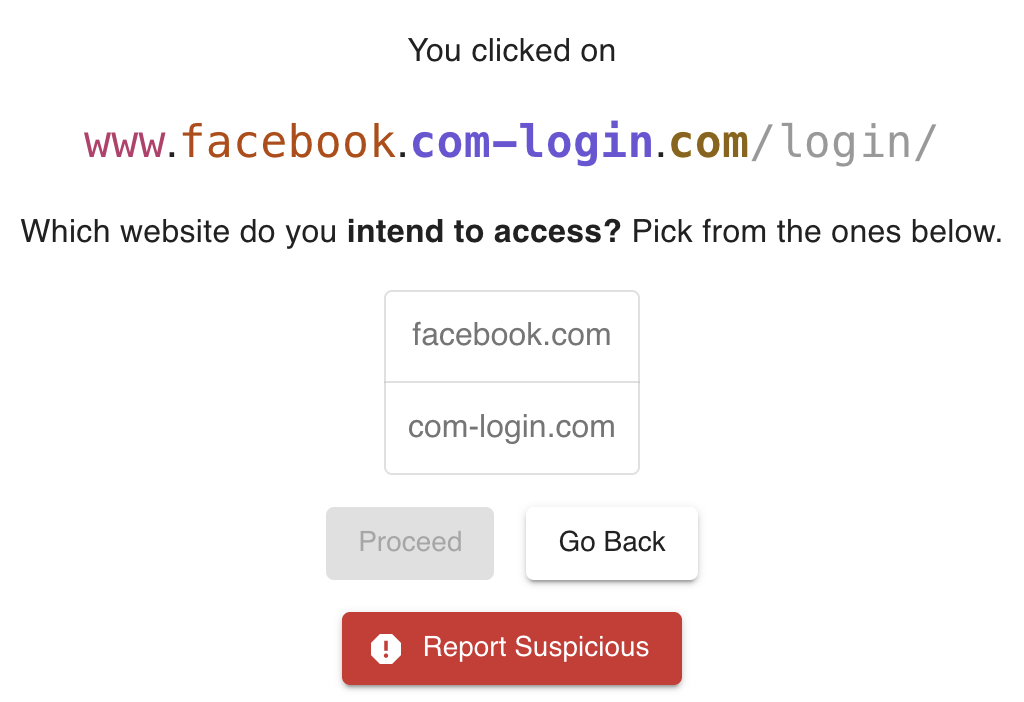}
        \caption{Clicking Task.}
        \label{fig:tasks.clicking}
    \end{subfigure}
    \begin{subfigure}{0.3\textwidth}
        \centering
        \includegraphics[width=\textwidth]{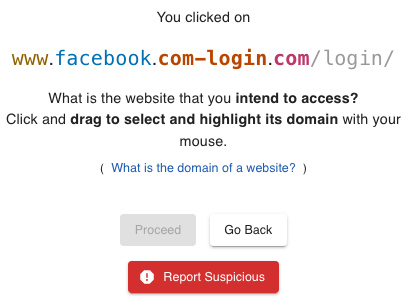}
        \caption{Highlighting Task.}
        \label{fig:tasks.highlighting}
    \end{subfigure}
    \begin{subfigure}{0.3\textwidth}
        \centering
        \includegraphics[width=\textwidth]{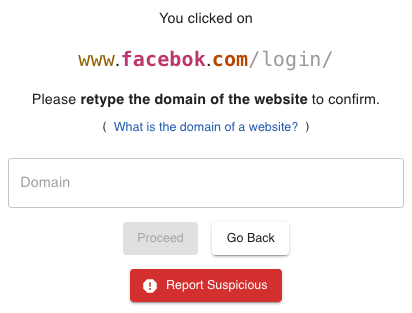}
        \caption{Typing Task.}
        \label{fig:tasks.typing}
    \end{subfigure}
    \caption{Three selected tasks for active URL inspection after clicking on a link.}
    \label{fig:tasks}
\end{figure*}

\subsection{URL Tasks}
\label{sec:design.tasks}

We selected three tasks based on three basic human-computer interaction (HCI) actions: clicking, 
dragging with the mouse, and typing. These are similar to most common CAPTCHA interactions~\cite{searles2023empirical}.
These tasks, shown in Figure~\ref{fig:tasks}, were designed so that performing them would force the user to re-read the URL and help them understand 
where it leads. This is in contrast with prior 
approaches~\cite{volkamer2017user,lin2011does,althobaiti2021don,petelka2019put}.
The tasks require the user to identify the domain portion of the URL. To do so, the user needs to (i) understand what a 
domain name is and how it identifies a specific website within a URL. However, as introduced, they also need to (ii) know the domain of the intended website: we discuss how users' knowledge affects the three selected tasks below; Section~\ref{sec:discussion} provides further details. 

\paragraph{Clicking Task}
Asking users to click on the domain itself~\cite{volkamer2017user} might lead to a simple form of habituation -- the domain would be 
presented roughly in the middle of a URL and users might click on it without paying attention.
Instead, our clicking task involves subdomains: we list the 
domain and subdomains in random order, \new{selected for example with heuristics on keywords}, and ask the user to click on the domain (Figure~\ref{fig:tasks.clicking}).

The main idea here is that this task should alert a user to a URL that contains a deceptive string in its subdomains or path.
In other words, a user would click on the domain they intend to visit. For example, given a choice between 
\texttt{example.\uline{com-login.com}} and \texttt{example.com} a user would click the latter.
This task can also help against deceptions within the domain itself, e.g., it can detect the presence of keywords within it with heuristics and propose the legitimate domain among the list.
Here, however, success would ultimately depend on a user's understanding that \texttt{example-login.com} is not the correct domain: we discuss this further in Section~\ref{sec:discussion}.
Finally, for typosquats, this task does not provide any specific help other than making a user re-read the URL.

\paragraph{Highlighting Task}
In this task, a user is asked to highlight the domain-name component of the URL (by clicking and dragging a mouse over it)
and then confirm by pressing a button, as shown in Figure~\ref{fig:tasks.highlighting}.

This task aims to capture a user's real intent for URLs that contain impersonations in subdomains, e.g., presented
with \texttt{example.\uline{com-login.com}}, a user would highlight \texttt{example.com} rather than \texttt{com-login.com}. 
It also aims to do the same for impersonations at the end of the domain (e.g., \texttt{\uline{login-example.com}}) 
and for ones contained in the URL fragment.
However, recognizing impersonations at the beginning of the domain requires a user to know that 
\texttt{\uline{example-login.com}} is not their intended (i.e., spoofed) domain, as they would (correctly, but without preventing the attack) highlight the whole domain otherwise.
Finally, for typosquats, this task helps users review the characters one at a time, potentially helping spot the deception.

Indeed, the modern URL structure presents a trade-off.
On the one hand, the relevant part of the URL for a given user might be the second-level subdomain, e.g., {\tt ``drive''} in 
\texttt{drive.google.com}, where the service name is embedded).
To accommodate this, the task must allow users to highlight subdomains.
On the other hand, this flexibility might lead users to highlight subdomains that contain impersonations 
of the service name, successfully completing the task while evading the mechanism's intended purpose.

\paragraph{Typing Task}
This task requires a user to re-type the domain in a text box, as shown in Figure~\ref{fig:tasks.typing}.
Its goal is to mitigate all types of impersonations since it allows the user to freely express their intent 
by entering the domain they intend to visit.
Simple techniques need to be employed to prevent users from copy-pasting the URL or dragging it into the textbox.

This task seems especially beneficial against subdomain and path 
impersonation. It is also effective against typosquats, since the user has to re-type the domain.
However, the same issue of knowledge of the correct domain remains for some types of URLs, e.g., when impersonations are at the beginning of the domain.
The main downside of this task is its user burden of having to type the (potentially long) domain character by character, 
This results in longer solving time, higher false positive rate, and increased user frustration.
\new{More advanced design, e.g., parsing natural language answers, could be considered to mitigate these issues.}

\subsubsection{Communicating Mistakes}

Different types of phishing URLs, tasks, and user errors require distinct feedback approaches.
One effective strategy is to alert users when the URL's domain does not match their response, asking if they still want to proceed.
Another approach is to provide more specific feedback for certain impersonations, such as \texttt{example-login.com}, by highlighting the discrepancy between the target URL's domain and the user's answer when showing the alert.
Meanwhile, for typing tasks, feedback might take the form of visually highlighting the difference between the user-typed URL and the actual URL to alert a user to typosquatting.

	\section{Experimental Setup}
	\label{sec:setup}
	To understand the effectiveness of three selected tasks, we conducted an online study where participants were asked
to play the role of an employee of a fictitious company and had to manage their virtual character's email inbox.
The inbox contained a mix of benign and phishing emails. A participant had to process benign emails and report 
phishing ones. To make the role-play as realistic as possible, we took advantage of participants' prior
knowledge of, and familiarity with, certain technologies. The study featured a familiar-looking email client (Figure~\ref{fig:platform.client}) and 
realistic-seeming emails.
Also, a participant could personalize their experience by selecting their preferred emails, services, and roles.
Details of the experimental setup are described below.

\subsection{Role-play Platform}

\paragraph{Task}
The goal for a participant was to manage their character's mailbox.
They were instructed to manage two types of emails: (1) for an email containing no links, they had to read it and mark 
it as completed, and (2) for an email with a link, a participant had to click that link to indicate that 
their character would visit the website and do what was asked.
If any email/link seemed suspicious, they were instructed to report it through a button in the email client.

Each participant had a time limit of 15 minutes to manage all emails.
For benign emails, the correct action was to either mark them as completed or click on the link, while, 
for phishing emails, the correct action was to report them.
While we are mainly interested in collecting data pertaining to emails with links, we introduced the additional 
task of asking participants to mark emails with no links as completed to make their 
experience more realistic and avoid priming them on the true nature of the study.

\paragraph{Steps}
As a setting for the email management task, we designed and developed a custom online platform.
First, participants gave their informed consent using a checkbox and button on a consent form that 
described the study as a role-play with the goal of testing a new user interface to an email client.
Next, participants filled out a pre-study questionnaire that collected demographic information and their 
familiarity with technology, phishing, and a set of popular online services: from document processing and 
sharing tools (Google Drive and Microsoft Sharepoint) to social media providers, payment platforms, and 
delivery services. These answers customized the role-play setting and 
the emails participants would receive, as discussed in Section~\ref{sec:setup.content}.

After the questionnaire, participants were introduced to their character: their role and responsibilities 
in the company, and basic information that their character would know, e.g., the format of corporate email addresses 
and names and URLs of various services used at the company, e.g., Google Drive or Microsoft Sharepoint.
Subsequently, on our custom browser-based email client mocked up to resemble Microsoft Outlook 
(Figure~\ref{fig:platform.client}), participants could manage the emails received by their character, 
besides reviewing information about the character and their role at the company.
Also, on the side of the screen, participants were always reminded of study instructions and what they had to do.
The email client featured a timer showing the remaining time to complete the task and a button to 
leave the study early if they wished to do so.

After managing the emails, participants were informed about the true nature of the study and received 
more information about phishing as well as the means to protect against it.
Finally, participants were presented with a post-study questionnaire which collected information about 
their study experience and the anti-phishing mechanisms that they encountered.

\subsection{Study Content}
\label{sec:setup.content}

\paragraph{Study emails}
We crafted a set of realistic-looking emails to be ``sent'' to the participants' characters according to their job
responsibilities. 
To enhance the study's realism and tap into participants' existing familiarity, we used real emails 
from well-known services for both legitimate and phishing emails.
By stripping away cues from the email content, we create a more realistic scenario where phishing e
mails are harder to detect, thus enabling a more accurate assessment of the proposed mechanism's effectiveness.
We created a total of 36 emails, as follows:
\begin{compactitem}
    \item 6 \textit{internal group emails}: benign text-only emails that set the context for the role-play and 
    familiarized participants with the names and email addresses of their co-workers.
    \item 9 legitimate and 9 phishing \textit{services emails}: mimicking those participants would receive 
    in their daily work routine from common services (e.g., comments on a Sharepoint document, or a FedEx tracking email).
    \item 6 legitimate and 6 phishing \textit{direct emails}: to test participants on more targeted attacks, such as spearphishing.
\end{compactitem}

\paragraph{Study URLs}
For each service used in the study, we created a set of six URLs: one legitimate URL and five phishing URLs, 
each representing a distinct type of phishing attack categorized in Section~\ref{sec:design.url}.
The legitimate URL was obtained directly from the actual service and, where necessary, included realistic path or 
query parameters, e.g., a Google Drive document URL featuring a path \texttt{/drive/folders/1t8FLJdJzDSOsMFYv} 
which incorporates the document ID.
Phishing URLs were constructed to be as similar as possible to the legitimate URL, the only difference being
the domain or the path component, Also, they were purposely designed to be confusing and hard to detect.
All URLs used in the study are shown in Appendix~\ref{sec:appendix.urls}.

\paragraph{Sampling}
\new{
Each participant received a total of 14 emails to manage: 11 legitimate and 3 phishing.
The exact emails served to each participant were customized according 
to the answers provided in the pre-study questionnaire: all {6 group emails}, 
{4 legitimate services emails} and {2 phishing services emails}, and {1 direct legitimate} and {1 direct phishing email}.}

\new{
Each benign email contained a link to its legitimate URL. For each phishing email, one of its 
five possible phishing patterns was picked at random.
Furthermore, for each URL, we randomly selected one of the three tasks they would be served upon clicking on the link 
(clicking, highlighting, or typing), with one exception: the clicking task was only served for phishing 
URLs that are {\bf not} typosquats, because they did not have any subdomains to generate the list of choices.}

\begin{figure*}[!t]
    \centering
    \begin{subfigure}{0.32\textwidth}
        \centering
        \includegraphics[width=0.9\textwidth]{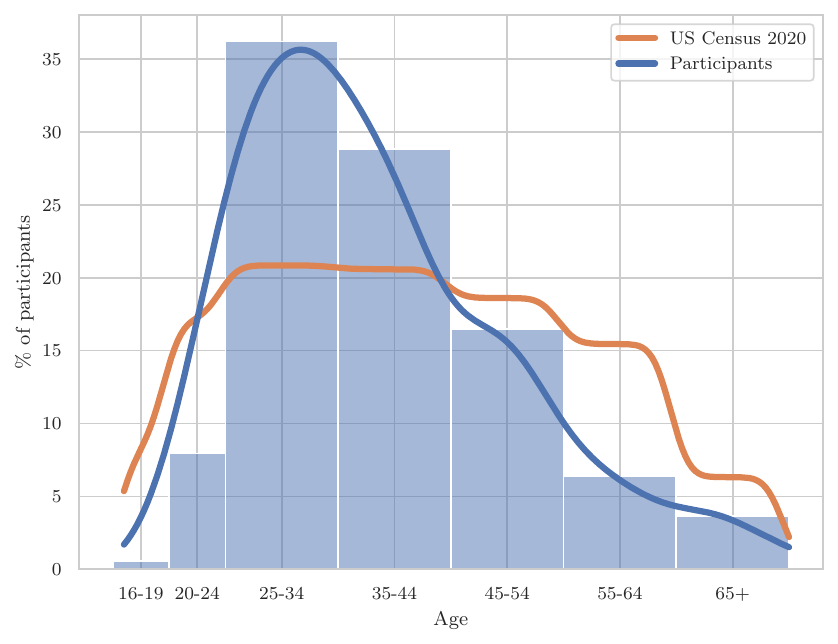}
        \caption{Age.}
        \label{fig:demographics.age}
    \end{subfigure}
    \begin{subfigure}{0.32\textwidth}
        \centering
        \includegraphics[width=0.9\textwidth]{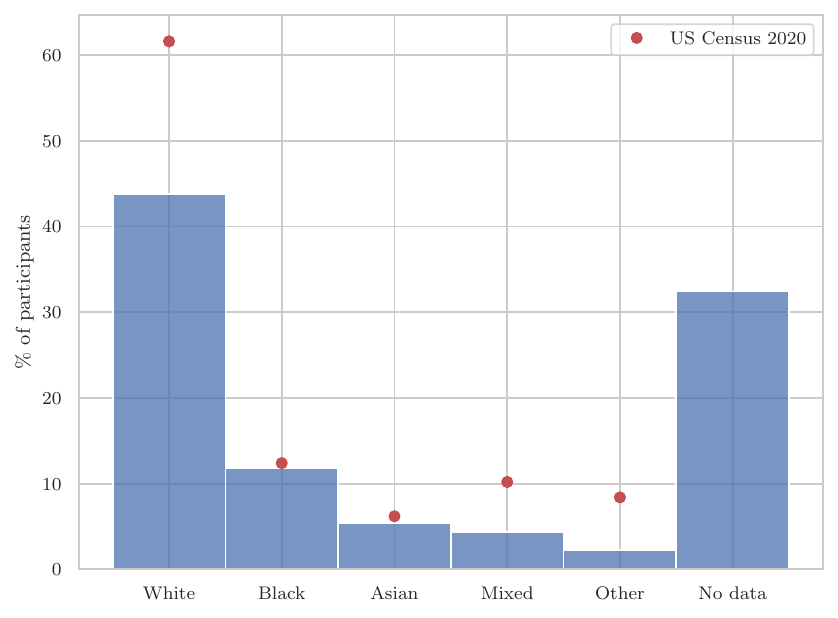}
        \caption{(Simplified) Ethnicity.}
        \label{fig:demographics.ethnicity}
    \end{subfigure}
    \begin{subfigure}{0.32\textwidth}
        \centering
        \includegraphics[width=0.9\textwidth]{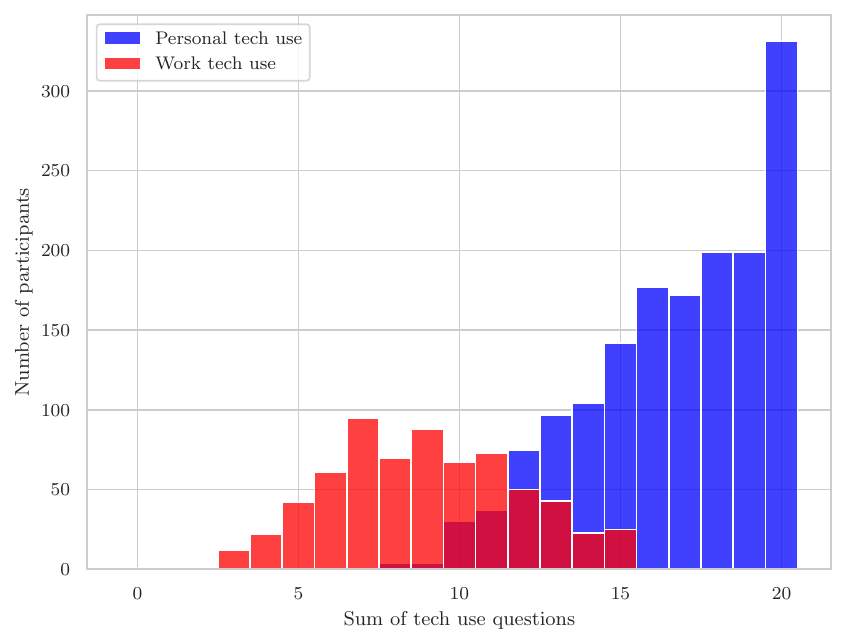}
        \caption{Tech Use Summary.}
        \label{fig:demographics.tech_use_summary}
    \end{subfigure}
    \caption{\textbf{Main demographics of the U.S. participants.} For age, we also report the distribution of the population 
    in employment in the U.S. in 2024; for ethnicity, the distribution of the general population from the U.S. 2020 census.}
    \label{fig:demographics}
\end{figure*}

\subsection{Experimental Groups}
We divided participants into 4 experimental groups according to the help and tools they received to manage the mailbox: 
a control group, two baseline approaches to compare against (see Appendix~\ref{sec:appendix.baselines} for details), 
and our tasks.

\noindent\textbf{Control}: participants did not receive any help and had to rely on their own knowledge 
and the email client interface.

\noindent\textbf{Passive} (baseline): after clicking on a link, participants were shown a warning page which 
presented the URL and asked them to confirm that they wish to visit it. This is a common approach used by several online services.

\noindent\textbf{Active} (baseline): participants were given with an activation task of dragging the pieces of the URL 
on which they just clicked to the center line and then confirm that they wish to navigate to that page.
This second baseline helps us decouple benefits of activation from those of intent checking: the task is 
designed to engage the user actively, though it cannot be solved incorrectly since the user has to notice 
whether it is a phishing URL while performing the task.
This latter aspect is only provided by our mechanism; thus, the comparison will help us isolate these two effects.

\noindent\textbf{Inspection tasks}: our novel tasks were served upon clicking a link in an email.

We decided to assign participants to the groups with an imbalance: roughly three times as many participants 
were assigned to our mechanism.
This is because we aim to study three different tasks, and thus wanted comparable group sizes for \textit{each} of them.

\subsection{Study Execution}
\paragraph{Participants}
We recruited participants for the main study on Prolific, a well-known and popular crowd-sourcing platform.
Participants has to be at least 18 years old, residing in the U.S., with English as their first language, 
a Prolific approval rate of at least 95\%, and at least 50 previous completed submissions on the platform.
Participants were paid to meet the highest minimum wage in the U.S., i.e., US\$~17.25/hour. 

The study did not employ attention or performance checks since they are not recommended 
and do not seem to improve data quality on the Prolific platform~\cite{tang2022replication,douglas2023data}.
Furthermore, previous work highlighted how attention checks can actively change participants' attention (rather 
than test it) and prime them to be more attentive since they are afraid of being tricked~\cite{hauser2015sa}, 
thus introducing an unacceptable bias. However, we excluded only \numExcluded{} participants due to mismatches 
between their answers in our questionnaires and the data they provided to Prolific or managing less than 
two emails before leaving the study. The median completion time for the study ranged from 9m 37s for the control 
group to 14m 02s for our tasks group; slightly more than 80\% of all participants completed the study within the estimated 20 
minutes. Furthermore, 97\% of participants fully filled out the pre-study questionnaire (18 or 19 answers), 
and 96\% the post-study questionnaire.

\paragraph{Demographics}
Key demographics of the participants can be seen in Figure~\ref{fig:demographics}.
The gender balance was: male -- \numMale, female -- \numFemale, and  other -- \numOther{}.
Figure~\ref{fig:demographics.age} shows that the participants base was skewed towards younger ages, 
in particular, 25-34 and 35-44, deviating slightly from the general population in employment in the U.S. 
and under-representing the 55-64 age group. Participants' ethnicity is shown in Figure~\ref{fig:demographics.ethnicity}: 
we observe a fairly balanced distribution compared to the general U.S. population.
Regarding education, 30\% had a high school diploma, 47\% a bachelors degree, and 17\% a masters degree.

Our initial questionnaire asked participants about the frequency of technology use in their private lives 
(computers, smartphones, instant messaging, and email) and on the job (computers for technical work, 
computers for non-technical work, e.g., data entry, and communication tools) on a scale from 1 to 5, 
where 1 is never, and 5 is all the time. Furthermore, we asked participants about their familiarity with 
email scams, the term ``phishing'', and whether they had received, or fallen for, any phishing emails in 
the past year, either in their personal lives or on the job. 

We report the sum of participants' answers related to the use of technology in their private lives 
(from 4 to 20) and in their jobs (from 3 to 15) in Figure~\ref{fig:demographics.tech_use_summary}. 
The participants are skewed towards being tech-savvy, with frequent use of electronic devices in their personal lives; 
there is a more even distribution in the use of technology on the job.
Participants reported a high perceived familiarity with email scams (75\% of participants 
reported a 4 or 5, mean 3.93) and the term phishing (similar numbers).
Participants frequently receive email scams in their personal mailboxes: 76.6\% received more than one in the last year. 
However, only 3.4\% reported falling for one, and 10.9\% almost falling for one.
There is more diversity related to phishing in the workplace: of the employed participants, 60\% experienced 
one or more phishing emails in the last three months, and 48\% received regular training in email security. 
This was similarly observed in previous studies using Prolific for security-related tasks~\cite{tang2022replication} 
that found greater technology use and more knowledge of technology among participants, as compared to the general population.

	\section{Results}
	\label{sec:results}
	\begin{table*}[t]
	\renewcommand{\arraystretch}{1.1}
	\footnotesize
	\centering
	\caption{\textbf{High-level performance.} \dotuline{Underlined values} are statistically significant compared to the control group, and \textbf{bold values} to the baselines. Conducted statistical analyses are Kruskal-Wallis test and post-hoc Dunn's test.}
	\label{tab:performance}
	\rowcolors{2}{white}{gray!10}
	\begin{tabularx}{\linewidth}{@{}ll|XXX|XXX|XXXX@{}}
		\toprule
		\textbf{Group} & \textbf{Size} & \multicolumn{6}{c|}{\textbf{Legitimate Emails} {\scriptsize \textit{(visiting is correct)}}} & \multicolumn{3}{c}{\textbf{Phishing Emails} {\scriptsize \textit{(reporting is correct)}}} \\
               &             & \multicolumn{3}{c|}{\textbf{Text-only}} & \multicolumn{3}{c|}{\textbf{With Link}} & Tot. & Visit & Report & Report \\
               &             & Tot. & Visit & Report & Tot. & Visit & Report & & & (Task) & (Mailbox) \\ \midrule

		Control & 248 & 744 & 94.8\% & 0.5\% & 1736 & 90.6\% & 8.6\% & 744 & 74.5\% & - & 24.9\% \\Passive & 240 & 720 & 98.2\% & 0.1\% & 1680 & 87.9\% & 11.4\% & 720 & \dotuline{57.2\%} & \dotuline{30.6\%} & 11.2\% \\Active & 269 & 807 & 95.4\% & 0.1\% & 1883 & 89.4\% & 10.5\% & 807 & \dotuline{61.0\%} & \dotuline{26.4\%} & 12.5\% \\Inspection Tasks & 816 & 2448 & 95.3\% & 0.4\% & 5712 & \dotuline{82.1\%} & 16.9\% & 2448 & \dotuline{\textbf{35.0\%}} & \dotuline{\textbf{51.9\%}} & 11.9\% \\
		\bottomrule
	\end{tabularx}
\end{table*}

\paragraph{\new{Methods}}
\new{Statistical significance was assessed everywhere by selecting a suitable statistical test according to normality criteria (Kruskal-Wallis unless otherwise specified). Differences were confirmed with a post-hoc Dunn's test. Every difference highlighted in the tables and text is significant with $p<0.05$.}

\subsection{Performance}

We first analyze the participants' performance in correctly identifying the study emails based on their belonging group. 
We report the results on the different legitimate and phishing emails divided per group in Table~\ref{tab:performance}.

\paragraph{Phishing detection}
We observe that all mechanisms outperform the control group in reporting and not falling for phishing emails: while the control group participants fell for 74\% of them, our tested baselines reduce this number to 57\% and 61\%, respectively.
However, our inspection tasks outperform all the other approaches by further reducing this number to 35\%, a difference that is statistically significant from all other groups, according to a Kruskal-Wallis test and a post-hoc Dunn's test.

\paragraph{False positives}
We also analyze the performance of participants in managing legitimate emails correctly.
We can see that, as expected, the performance on emails without links is similar across all groups and is very high, ranging from 95.0\% to 98.2\%.
For emails with links, the performance is slightly different across groups.
While the two baseline groups perform very similar to the control group (87\% to 89\%), participants receiving our tasks perform 5\%-7\% worse, as they become overly suspicious of some legitimate URLs and report slightly more legitimate emails.
\new{However, this difference is not statistically significant; some increase can be attributed to the roleplay scenario and to the heightened alertness after encountering phishing URLs for participants not in control.}

\begin{table}[t]
	\renewcommand{\arraystretch}{1.2}
	\footnotesize
	\centering
	\caption{\textbf{Phishing rates for different URL types:} \textsc{sub} with impersonation  in subdomain; \textsc{first} at the beginning of the domain; \textsc{last} at the end; \textsc{path} in the path; \textsc{squat} typosquats.
	\dotuline{Underlined values} are statistically significant compared to the control group, \textbf{bold} to either baseline, \textbf{*} to all baselines, according to a Kruskal-Wallis and post-hoc Dunn's test.}
	\label{tab:performance_per_url_type}
	\rowcolors{2}{white}{gray!10}
	\begin{tabularx}{\linewidth}{@{}X|XXXXX@{}}
		\toprule
		 & \textsc{sub} & \textsc{first} & \textsc{last} & \textsc{path} & \textsc{squat} \\ \midrule

		Control & 75.7\% & 76.8\% & 72.0\% & 72.0\% & 79.2\% \\
Passive & 63.4\% & \dotuline{58.3\%} & 56.5\% & 55.5\% & \dotuline{57.8\%} \\
Active & \dotuline{53.5\%} & 70.6\% & 64.0\% & 59.2\% & \dotuline{58.6\%} \\
\midrule
Click & \dotuline{\textbf{31.2\%*}} & \dotuline{\textbf{46.4\%}} & \dotuline{\textbf{39.3\%*}} & \dotuline{\textbf{36.5\%*}} & - \\
Highlight & \dotuline{\textbf{35.1\%*}} & \dotuline{\textbf{44.4\%}} & \dotuline{\textbf{44.4\%}} & \dotuline{\textbf{41.7\%}} & \dotuline{47.2\%} \\
Type & \dotuline{\textbf{27.5\%*}} & \dotuline{\textbf{42.9\%}} & \dotuline{\textbf{41.2\%}} & \dotuline{\textbf{41.0\%}} & \dotuline{\textbf{17.1\%*}} \\

		\bottomrule
	\end{tabularx}
\end{table}

\subsection{Different URL Types}

We further analyze the participants' performance on different types of phishing URLs.
The results are reported in Table~\ref{tab:performance_per_url_type}.

We observe that the baseline approaches only show slight improvements over the control group, with the passive task showing statistically significant differences only for impersonations at the beginning of the domain and typosquats and the active one only for subdomain impersonation and typosquats.
Meanwhile, the proposed tasks show a statistically significant improvement over the control group for every type of task and phishing URL.
We can also see that all proposed tasks present statistically significant improvements over the baselines for all types of phishing URLs, except highlighting typosquat URLs which showed a minor 11\% improvement over the baseline.

All of the proposed tasks are highly effective, presenting improvements over even the best performing baseline with 10\% to 40\% more reported emails and less success of phishing URLs.
In particular, the best performing tasks for each type of phishing URL provide significant improvements: for impersonations in subdomains, 26\% fewer phishes are successful; for the beginning of the domain, 15\% less; for the end of the domain, 17\% less; for the path, 19\% less; and for typosquats, 40\% less.
Especially notable is the improvement for typosquats, which were among the hardest to detect for participants in the control and baseline groups, but our mechanism achieves a three-fold improvement.

These results empirically confirm the discussion of the different effectiveness of tasks for various types of URLs that we presented in Section~\ref{sec:design.tasks}.
In particular, highlighting helps less for typosquats as it is limited to slowing down the user and attracting attention to the single characters.
Instead, clicking on the intended domain from a list proves highly effective as it allows users to clearly realize if their intention mismatches the URL, by picking the desired (correct) one from the list.
Finally, as expected, typing was highly effective against typosquats, as the user is unlikely to retype the URL correctly (i.e., with the wrong character) and thus will be notified of the mistake.

\begin{table}[]
	\renewcommand{\arraystretch}{1.2}
	\footnotesize
	\centering
	\caption{\textbf{Typosquat URLs:} phishing rates per group.}
	\label{tab:performance_per_typosquat}
	\rowcolors{2}{white}{gray!10}
	\begin{tabularx}{\linewidth}{@{}l|XXX|XX@{}}
		\toprule
		& \textbf{Control} & \textbf{Passive} & \textbf{Active} & \textbf{Highlight} & \textbf{Type} \\ \midrule

		Addition & & & & & \\$\quad$\texttt{fed-ex} & 64.7\% & 60.0\% & 70.0\% & 14.3\% & 25.4\% \\
Deletion & & & & & \\$\quad$\texttt{facebok} & 75.0\% & 58.1\% & 56.5\% & 66.7\% & 13.0\% \\
Substitution & & & & & \\$\quad$\texttt{sharep0int} & 84.6\% & 33.3\% & 43.8\% & 75.0\% & 24.0\% \\
$\quad$\texttt{googie} & 91.3\% & 63.6\% & 53.6\% & 22.2\% & 10.0\% \\
$\quad$\texttt{linkedln} & 87.5\% & 68.3\% & 80.8\% & 63.6\% & 20.3\% \\
$\quad$\texttt{paypai} & 61.1\% & 40.0\% & 30.0\% & 33.3\% & 6.7\% \\
Swap & & & & & \\$\quad$\texttt{mircosoft} & 88.2\% & 50.0\% & 68.4\% & 44.4\% & 13.0\% \\

		\bottomrule
	\end{tabularx}
\end{table}

\subsubsection{Typosquat URLs}
We further turn our attention to typosquat URLs, as it is interesting to compare the performance of all tasks on different types of typosquats: character addition, deletion, substitution, and transposition.
We report all typosquat URLs and the phishing rates for each group and task in Table~\ref{tab:performance_per_typosquat}.

We observe that typosquats are especially difficult for participants in the control group, who can only rely on the small tooltip offered by browsers when hovering over the link:
60\% fell for the PayPal typosquat, and a staggering 90\% for the Microsoft and Google ones.
This is understandable: character swaps are hard to detect because we tend to reorder them in our mind while substituting an \texttt{l} with an \texttt{i} can even be confused with a speckle of dust on one's screen.

Meanwhile, even the baselines show improvements over the control group; showing the URL with a bigger, monospaced font helped participants---with one notable difference: FedEx, where our baselines performed as poorly as the control group.
We attribute this to the fact that the FedEx typosquat requires knowledge of the correct domain due to simply adding a dash, while all other typosquats presented misspelling errors.
Yet, even for this difficult URL, the proposed tasks still show a marked improvement, reducing the falling rates from 65\% down to 14\%.
For all other URLs, all baseline approaches are shown to help participants.
This trend is also seen in the proposed tasks, where we observe the numbers improving, reducing the falling rates five to tenfold compared to the control group, especially on the more difficult Microsoft and Google typosquats, where our typing task decreased the falling rates from 88\% and 91\% to 13\% and 10\%, respectively.

{
\begin{table*}[t]
	\renewcommand{\arraystretch}{1.2}
	\footnotesize
	\centering
	\caption{\textbf{Participant actions per phase:} while solving the task and after solving it incorrectly.}
	\label{tab:mechanism_perf_per_phase}
	\rowcolors{2}{white}{gray!10}
	\begin{tabularx}{\linewidth}{@{}ll|llXXX|llXX@{}}
		\toprule
		&  & \multicolumn{5}{c}{\textbf{Task Solving}} & \multicolumn{4}{c}{\textbf{Mistake Page}} \\
		&  & Emails & Solved & Wrong & Report & Back~(report) & Emails & Confirm & Report & Back (report) \\ \midrule

		 & Legitimate & - & - & - & - & - (-) & - & - & - & - (-) \\
\multirow{-2}{*}{Click} & Phishing & 658 & 24.9\% & 36.6\% & 28.6\% & 11.4\% (2.6\%) & 241 & 36.1\% & 46.1\% & 20.7\% (6.2\%) \\
 & Legitimate & 2812 & 54.6\% & 29.5\% & 11.1\% & 13.8\% (2.1\%) & 829 & 90.5\% & 4.6\% & 8.1\% (0.7\%) \\
\multirow{-2}{*}{Highlight} & Phishing & 663 & 23.8\% & 30.6\% & 33.3\% & 11.9\% (2.3\%) & 203 & 57.1\% & 34.5\% & 13.8\% (1.0\%) \\
 & Legitimate & 2846 & 67.7\% & 19.4\% & 8.0\% & 10.1\% (1.1\%) & 551 & 87.5\% & 6.7\% & 7.8\% (0.4\%) \\
\multirow{-2}{*}{Type} & Phishing & 1100 & 16.3\% & 30.8\% & 43.2\% & 12.4\% (3.4\%) & 339 & 45.4\% & 33.0\% & 25.4\% (3.8\%) \\

		\bottomrule
\end{tabularx}
\end{table*}
}

\subsection{Which Components Are Beneficial?}

The next question we address is which components are most beneficial to participants.
Recall that our mechanism allows different interactions: when presented with a URL and task, a user can choose to solve it, report it and its email, or return to the email to inspect it again.
Users who choose to solve the challenge might do it incorrectly, and are presented a failure page displaying the URL and their solution, and allowing them to proceed anyway, report, or go back to their mailbox.
We postulate that all these interactions can contribute differently to the participants' performance, and thus raise the following question: how do each of these components contribute to our mechanism's effectiveness?
To do so, we report a breakdown of all interactions divided into legitimate and phishing URLs in Table~\ref{tab:mechanism_perf_per_phase}.

We observe that for legitimate URLs, users generally solve our tasks correctly (55\% to 68\%), and overwhelmingly (87\% to 90\%) confirm that they still want to proceed when they solved them incorrectly.
This is not the case for phishing URLs: while a minority of users (16\% to 25\%) solve the tasks correctly, most (28\% to 43\%) report the email---comparably or better than all the baselines. 
As there are more reports than in the passive and active baselines, the task design itself is effective in helping participants.

For phishing URLs, the remaining (31\% to 37\%) tasks are solved incorrectly, triggering communication of the mistake: on this second phase, roughly half of the participants' decisions are the ``correct'' behavior (reporting or going back), thus showing that triggering the mistake helps participants even further.
However, we observed that only a minority of the participants who went back and inspected the email again ended up reporting it from the mailbox; most of them proceeded to go through the task again.

We summarize the analysis by claiming that the components of the proposed tasks are effective in helping participants at all stages: while re-reading the URL, while solving the task, and after being communicated their mistake when solving the challenge.

\begin{figure}[t]
	\centering
	\begin{subfigure}{\columnwidth}
		\centering
		\includegraphics[width=.8\columnwidth]{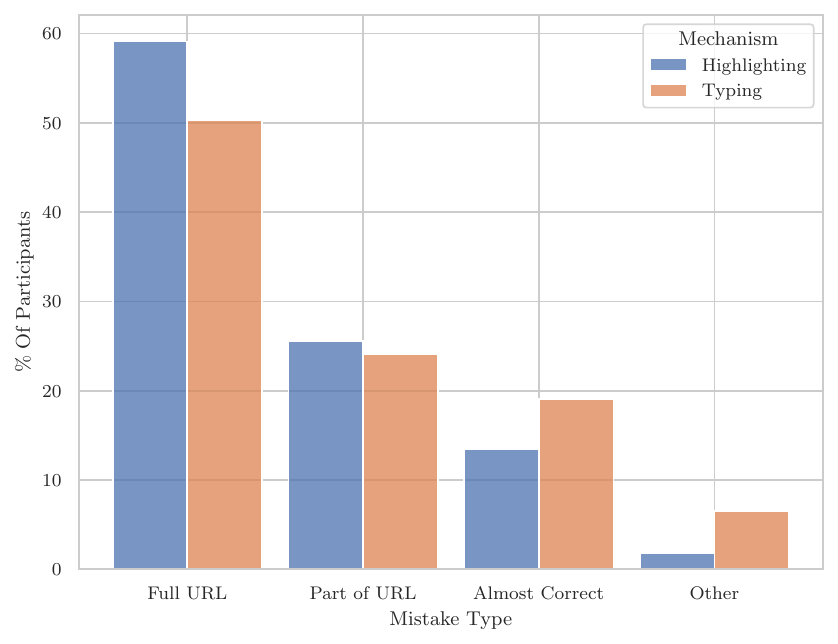}
		\caption{Legitimate URLs.}
		\label{fig:mistakes_types.legitimate}
	\end{subfigure}
	\begin{subfigure}{\columnwidth}
		\centering
		\includegraphics[width=.8\columnwidth]{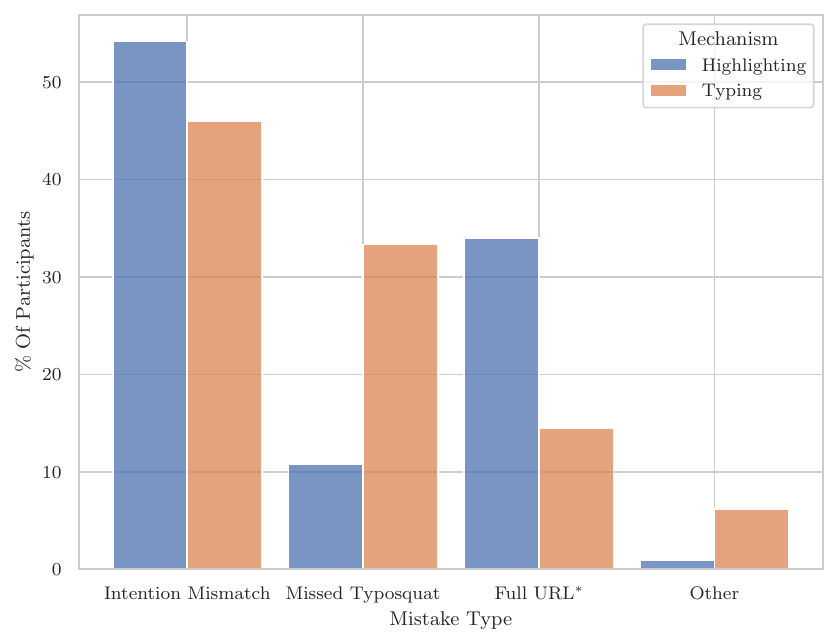}
		\caption{Phishing URLs.}
		\label{fig:mistakes_types.phishing}
	\end{subfigure}
	\caption{\textbf{Types of mistakes} made while solving the tasks.}
	\label{fig:mistake_types}
\end{figure}

\subsubsection{Types of Mistakes}

The final question related to the usefulness of our mechanism is about the mistakes participants make when solving the tasks.
We are interested in observing what mistakes are more common for legitimate and phishing URLs---the former to analyze how the participants misunderstand URLs and the tasks; the latter to understand whether the tasks can trigger mistakes related to mismatches of the participants' intentions with the URLs.
We manually go through all the mistakes and classify them into different types, reported in Figure~\ref{fig:mistake_types}.

Figure~\ref{fig:mistakes_types.legitimate} shows the types of mistakes on legitimate URLs: we observe that the most common mistake is to highlight or retype the full URL or parts of it instead of the domain, showing that participants struggled with the concept of domains (despite our interface included a button that explains just that, located next to the tasks).
A minority of participants also made minor mistakes in the solution, e.g., mistyping a character, missing a dot, or highlighting a few extra characters.

Figure~\ref{fig:mistakes_types.phishing} shows which mistakes were made on phishing URLs.
We observe that most of the mistakes users made relate to confusion due to the phishing URLs: the most common mistake (45\% to 50\%) is highlighting or typing the domain of the impersonated brand instead of the domain, e.g., \texttt{paypal.com} instead of \texttt{com-login.com} for the phishing URL \texttt{paypal.com-login.com}.
This highlights how the proposed tasks can help by spelling out the mismatch between the participant's intentions and the URL they are about to visit.
The other main source of mistakes (more than 30\% for typing) is not noticing typosquats and typing the URL correctly instead.
The least desirable mistake, i.e., highlighting or typing the full URL which would lead participants to the phishing website, only happened 33\% of the time for highlighting and 15\% for typing, proving that the main mistakes that our tasks triggered were \textit{helpful} ones, potentially leading participants to notice the scams.

\begin{figure}[t]
	\centering
	\includegraphics[width=.75\columnwidth]{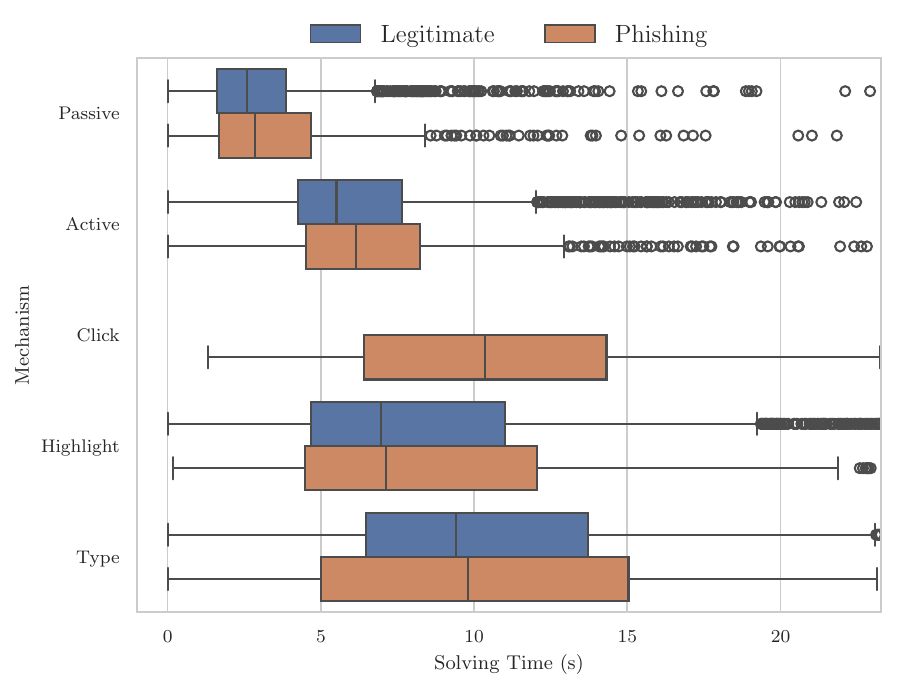}
	\caption{\textbf{Solving time per task} by email type.}
	\label{fig:solving_time}
\end{figure}

\subsection{Solving Time}

We analyze the solving times of the participants in solving the tasks, reported in Figure~\ref{fig:solving_time}.
While the passive baseline was solved quickly (median time: 3 sec), the active baseline and our tasks took longer: the fastest tasks to solve were the active one and highlighting (median time: 6-7 sec), while the typing task took the longest (median time: 10 sec).
However, all tasks exhibit large variances in solving times, similar to what is observed for CAPTCHAs~\cite{searles2023empirical}.
Moreover, we observe that for all tasks, including the baselines, there are almost no differences in the median solving time for legitimate and phishing URLs.
We further analyze the solving times per demographic to see whether any of the recorded participants' attributes have any influence.
We observed that age impacts the passive, active, highlight, and typing solving times; education the passive, active, clicking, and typing times; and technology use the passive and clicking times.
These differences are statistically significant but not very large---we report them in Appendix~\ref{sec:appendix.demographics_time}.

\subsection{Effects of Demographics}
\label{sec:results.demographics}

We analyze the effect of participants' demographics on their accuracy in managing both legitimate and phishing emails for all groups.
For this, we consider the participants' age, education, and reported technology use in their personal lives and work (aggregated by summing the answers for different devices and divided into low, medium, and high).

We observed only minor differences overall.
For the control group, no demographic attribute significantly affected either accuracy.
The same holds for our active baseline.
For the passive baseline, age and education affected phishing email accuracy only, with older participants performing worse and participants with higher education performing better.
For our tasks, we observed a similar statistically significant effect of age on legitimate email accuracy and the frequency of use of technology in private life on phishing email accuracy.
Curiously, the difference is not between the most and least frequent users but between the most and average frequent ones, with the latter performing worse---having observed no difference with the lowest use group suggests that potentially, less confident participants were more alert.
We report the detailed results in Appendix~\ref{sec:appendix.demographics_accuracy}.

\subsection{User Perception}
We now analyze our participants' perceptions of our mechanism.
We administered a post-study questionnaire asking participants about their experience with the study with Likert-scale questions (reported in Appendix~\ref{sec:appendix.post_questionnaire}).
We report the distribution of users' answers to the high-level questions related to our mechanism in Figure~\ref{fig:postq_mechanism_general}: we observe that most participants found our challenges helpful (Q1) and useful (Q2) in spotting phishing URLs.
Further appreciated was the presentation of the mechanism, which was found clear (Q9), with appreciated features such as coloring the URL (Q7), that our task made simpler to read and understand (Q8).
The tutorial (Q5, Q6) was well received, giving us confidence that the participants understood how to solve our tasks for the study.
The response is more mixed for clarity in highlighting mistakes (Q3), indicating potential for improvement.
Finally, while the mechanism did not feel obtrusive (Q4), we have to note that participants only solved our tasks a handful of times in a short time span, with the goal of getting a study reward.

\begin{figure*}[t]
	\centering
	\begin{subfigure}[c]{0.65\textwidth}
		\centering
		\includegraphics[width=\textwidth]{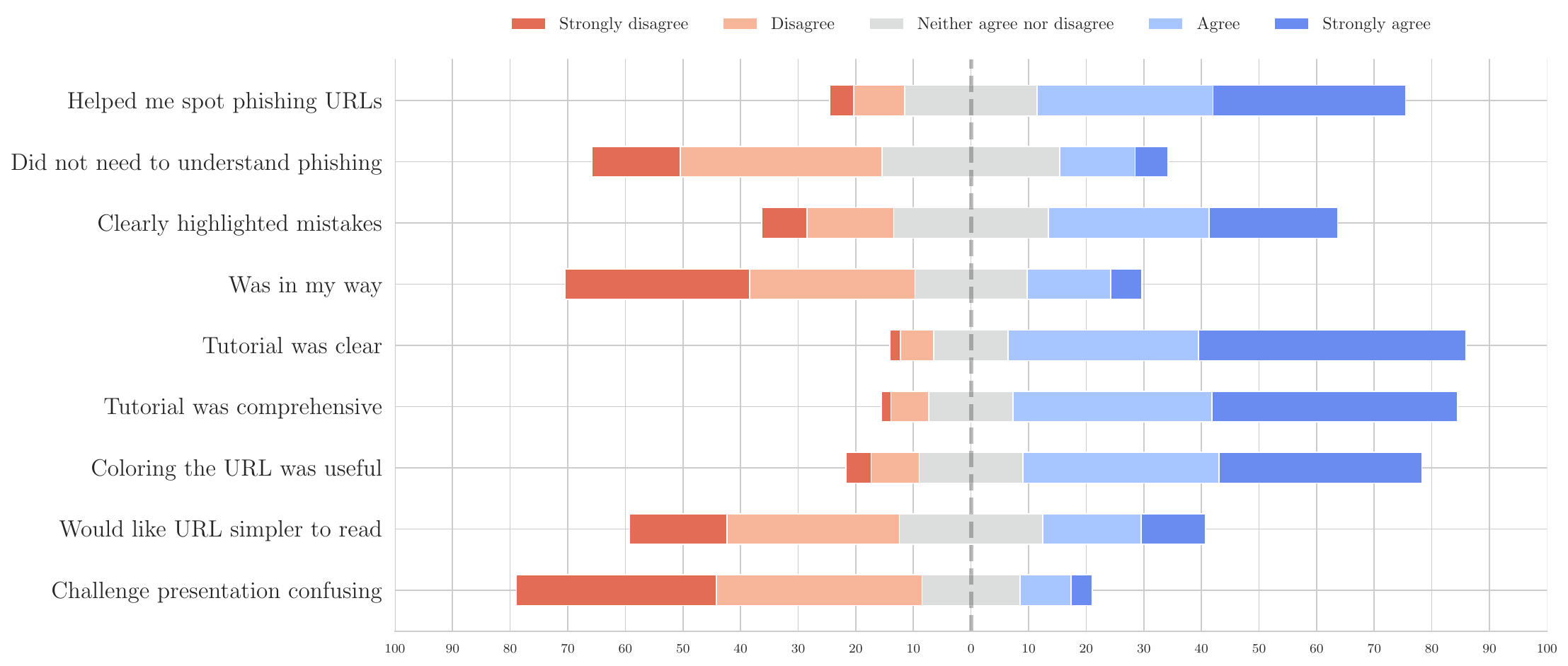}
		\caption{Common questions.}
		\label{fig:postq_mechanism_general}
	\end{subfigure}
	\begin{subfigure}[c]{0.33\textwidth}
		\centering
		\includegraphics[width=\columnwidth]{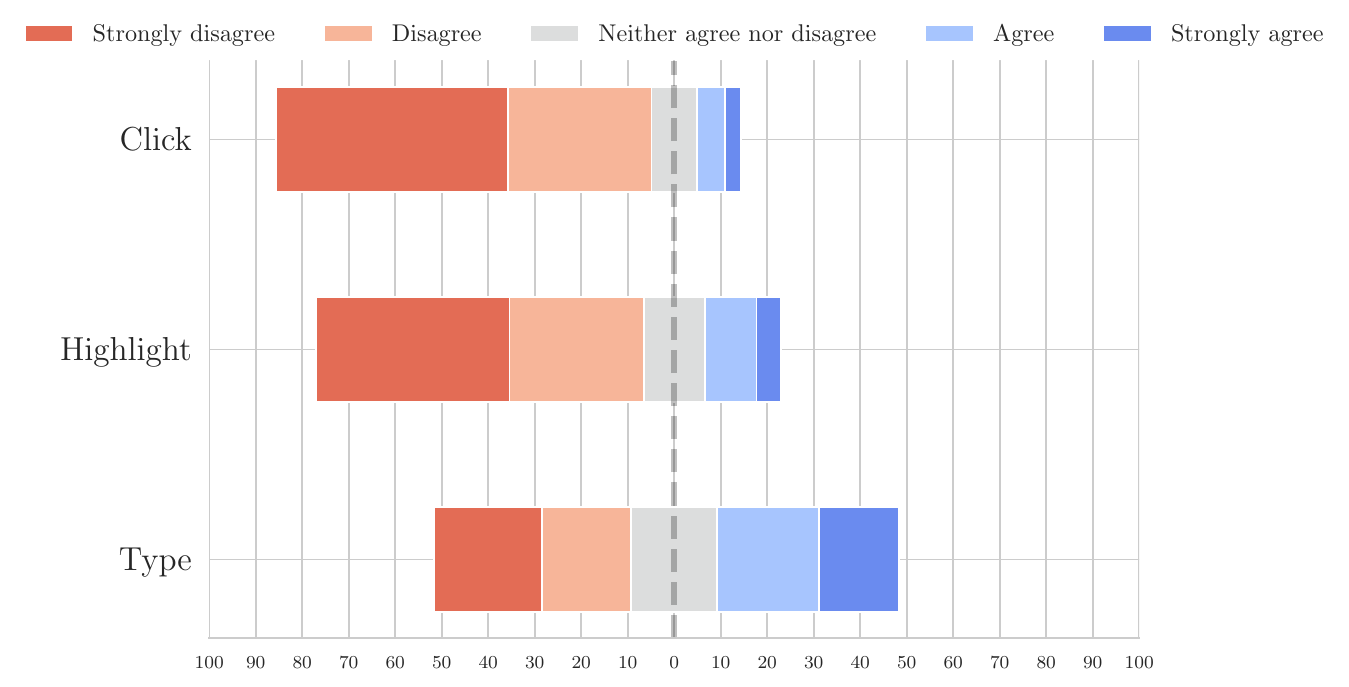}
	\caption{Q2: {\textit{``I found the task annoying''}}.}
	\label{fig:post_mechanisms_q1}
	\end{subfigure}
	\caption{\textbf{Post-study questionnaire answers} for general aspects of our tasks, and for Q2 for each task.}
	\label{fig:postq_mechanism_general}
\end{figure*}

Finally, we analyze the participants' answers to task-specific questions: whether participants found them (i) useful, (ii) annoying, and (iii) difficult.
Here, the vast majority of participants found all our tasks useful and not difficult; however, the typing task was found more annoying than the other two, as we report in Figure~\ref{fig:post_mechanisms_q1}.
This result is similar to observations made on CAPTCHAs, where the ones who require more effort are also the most disliked~\cite{searles2023empirical}, suggesting trade-offs between task efficacy and user experience.

\section{\new{More Complex and Frequent Usage}}

\new{We further investigate the performance of our mechanism in more challenging scenarios.
In particular, while the recommended usage of our proposed countermeasure is sporadic, we are interested in understanding how user performance and perception change when exposed frequently, especially whether frequently encountering our tasks on legitimate URLs would, e.g., lead to habituation, increased false positives, or annoyance.
Furthermore, we are interested in testing the performance of our tasks on more challenging URLs, such as those with more subdomains, or less commonly known ones.}

\new{To do so, we conducted a follow-up study with 500 participants from the US, where we doubled the number of emails from 13 to 25 (18 legitimate and 7 phishing), and the time to complete the study to 30 minutes. Furthermore, we added 4 more challenging URLs, e.g., with more subdomains and less intuitive names, such as cloud services like \texttt{azure.com}.
We report these new URLs in Appendix~\ref{sec:appendix.urls}.
To allow for better comparisons, the demographic distribution of participants was the same of the main study. Participants were randomly assigned to one group between control (100 participants), passive baseline (100), and inspection tasks (300).}

\subsection{\new{Performance Under Longer Exposure}}

\new{We first compare the performance of participants in the follow-up study on the emails and URLs that were also present in the main study.
We report the results in Table~\ref{tab:performance_followup}.
We can observe that, while in the longer study all groups performed slightly worse, our mechanisms still outperformed both control and the baseline approach, and significantly helped participants with a high improvement of 35\% less successful phishing compared to control.}

\new{We observe the same increase in false positives for legitimate emails compared to the control group, however, this did not get larger despite the 3x increase in the number of legitimate emails to manage.
Furthermore, the baseline group observed a similar increase in false positives.
Finally, time overhead for each task remained the same as in the main study.
Combined with the fact that most of these URLs would be allowlisted in a corporate setting and the sporadic nature of our countermeasure, these results suggest a tolerable increase in false positives.}

\begin{table}[t]
	\renewcommand{\arraystretch}{1.1}
	\footnotesize
	\centering
	\caption{\new{\textbf{Main and follow-up studies:} results comparison.}}
	\label{tab:performance_followup}
	\rowcolors{2}{white}{gray!10}
	\begin{tabularx}{\linewidth}{l|XX|XX}
		\toprule
		\textbf{Group} & \multicolumn{2}{c}{\textbf{Legitimate Emails Managed}} & \multicolumn{2}{c}{\textbf{Phishing Victimization}} \\
		& Main & \mbox{Follow-up} & Main & \mbox{Follow-up} \\ \midrule

		Control & 90.6\% & 93.5\% & 74.5\% & 83.3\% \\
		Passive & 87.9\% & 87.8\% & 57.2\% & 69.1\% \\
		Inspection & 82.1\% & 82.9\% & 35.0\% & 48.8\% \\

		\bottomrule
	\end{tabularx}
\end{table}

\begin{figure}
	\centering
	\includegraphics[width=\columnwidth]{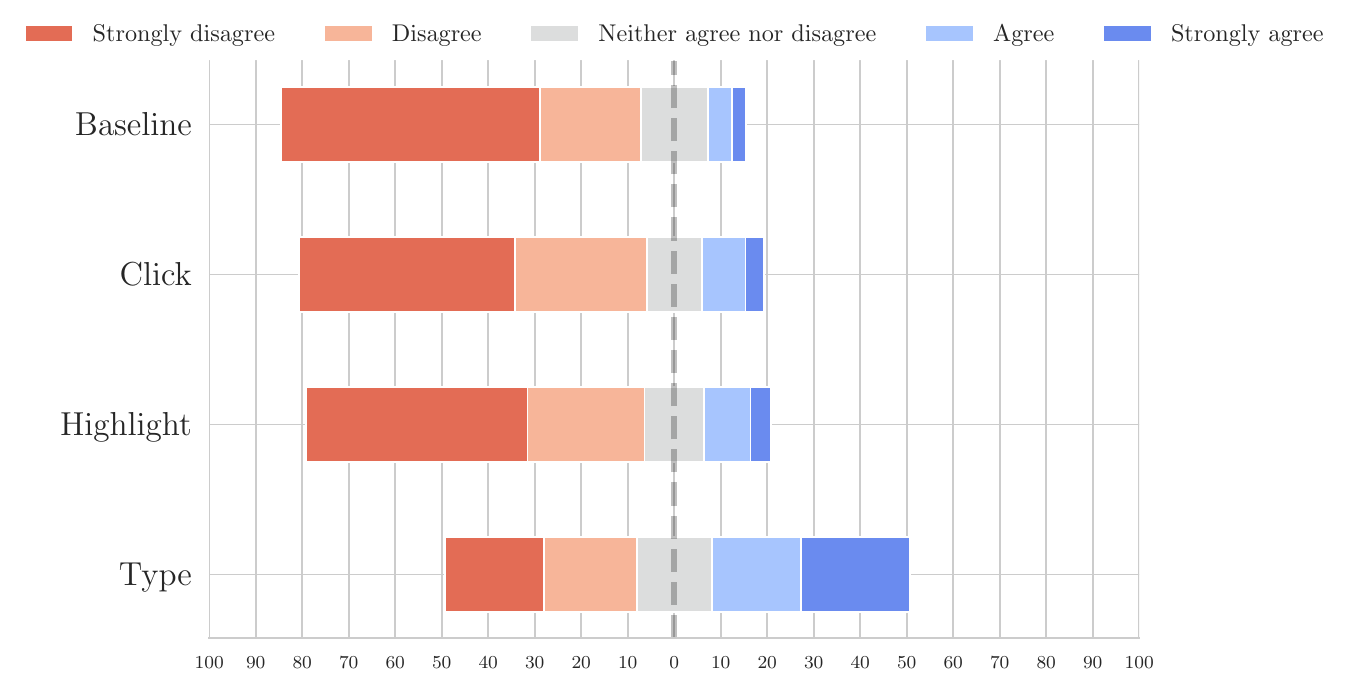}
	\caption{\new{\textbf{Longer exposure Q2:} \textit{``I found the task annoying''}.}}
	\label{fig:mr_postq_mechanism_q1}
\end{figure}

\new{To further assess the impact of longer exposure, we compared the answers of participants to the post-study questionnaire, especially regarding their perceived annoyance towards the tasks.
We show the result to the question \textit{``I found the task annoying''} in Figure~\ref{fig:mr_postq_mechanism_q1}. Compared to the main study, we observed similar annoyance, with the clicking and highlighting tasks being perceived as annoying as simply re-reading the URL, and the typing task being perceived as more annoying, but only slightly increasingly so compared to the main study.}

\subsection{\new{More Complex URLs}}
\label{sec:results.more_complex_urls}

\new{We now analyze in more detail the performance of participants on the 4 more challenging legitimate and phishing URLs we introduced (see Appendix~\ref{sec:appendix.urls}).
Both the legitimate and phishing URLs featured more subdomains and less intuitive top domains, such as cloud services, or legitimate but less known services such as \texttt{spreadsheet0.google.com}.}

\new{The performance on legitimate URLs (and thus the false positives rate) was similar to the one on simpler URLs on both the main and follow-up studies for the control group and our mechanism.
However, interestingly, the passive baseline group showed a significant decrease in performance (from 88\% to 82\%), highlighting that our tasks helped more than simply presenting the URLs.
For phishing URLs, we again observe our countermeasure helping greatly in reducing the falling rates with a smaller but statistically significant improvement (25\% less phishing emails clicked) over both control and the baseline group.
Furthermore, we analyzed the performance per mechanism and observed the same positive effects as in the main study, especially for the clicking task as in these URLs participants had to choose among 3 instead of 2 options.}

\paragraph{\new{Unknown and wrong URL}}
\new{Finally, we analyze the \textit{googleusercontent} URL (see Appendix \ref{sec:appendix.urls}), that we added both in a phishing email and a legitimate one. 
It represents a special case because it is the only instance of a legitimate email containing a suspicious URL (it was not in the list of legitimate domains presented to the participant).
Further, it is especially difficult containing multiple subdomains, an IP address, and a less known cloud services domain. 
Thus, we wanted to see whether participants would also report the email coming from a legitimate source.
Both our mechanism and the passive baseline helped participants to not fall for the phishing email and report it (improving an already very high 60\% to 82\% and 87\%, respectively).
However, interestingly, neither helped participants question the legitimate one as report rates were similar across all groups with non significant differences, as participants most likely used cues from the email to decide.
This scenario simulated a genuine mistake by a colleague, but could also represent a \textit{business email compromise} scenario, where an attacker has taken over a legitimate email account and sends phishing emails to the victim's contacts, and highlights its danger.}

	\section{Different Languages and Scripts}
	\label{sec:languages}
	Finally, we investigate the impact of different languages and scripts on the effectiveness of our tasks.
Our main study featured English-centric URLs in the Latin script---the most common script for URLs.
We ask ourselves: do phishing susceptibility and the usefulness of our tasks differ for non-native English speakers and for non-native Latin script readers?
Is reviewing URLs in a different script more challenging, and do the interactions we propose help users more or less in these cases?
To answer these questions, we had our study fully localized by native speakers, with the exception of the URLs, and ran it on 300 German participants (non-native English speakers but native Latin script readers) and 300 Japanese participants (non-native English speakers and non-native Latin script readers).
We picked these large countries as they have access to the same Internet services as the initial U.S.-based study (e.g., Google, PayPal) and have good availability of crowdsourcing platforms: Prolific for German speakers and Lancers for Japanese speakers.
We now present in the following a comparison with the U.S. study.

\begin{table}[t]
	\renewcommand{\arraystretch}{1.1}
	\footnotesize
	\centering
	\caption{\textbf{Different scripts:} phishing rates per group.}
	\label{tab:performance_languages}
	\rowcolors{2}{white}{gray!10}
	\begin{tabularx}{\linewidth}{@{}l|XXX@{}}
		\toprule
		& U.S. & Germany & Japan \\
		\midrule
		\textbf{Control} & 74.5\% & 53.2\% & 82.2\% \\
		\textbf{Passive} & 57.2\% & 38.4\% & 60.7\% \\
		\textbf{Active} & 61.0\% & 43.4\% & 67.3\% \\
		\textbf{Inspection Tasks} & 35.0\% & 25.9\% & 57.1\% \\
		\bottomrule
	\end{tabularx}
\end{table}

\paragraph{Demographics}
We briefly report on the demographics of the German and Japanese participants.
The German participants were similarly distributed to the U.S. participants in terms of all recorded variables---importantly, age, education, reported level of technology usage, and phishing awareness.
The Japanese participants instead had a very different age distribution: participants aged 31-40 were 27\%, 41-50 were 40.5\%, and 51-60 were 19.1\%---a significantly older population than the U.S. and German studies.
They also had more varied (and lower) reported levels of technology usage.

\paragraph{Study performance}
We report the high-level results for the German and Japanese studies in Table~\ref{tab:performance_languages}.
For German participants, we observed an overall lower phishing susceptibility compared to the U.S. and Japanese participants, including, most notably, the control group. 
However, the results are in line with the U.S. study: while both baselines helped participants, the proposed mechanism was the most effective, halving the phishing rate of the control group.
Results for the Japanese participants are more nuanced: while their control group and baselines exhibit slightly more susceptibility than their U.S. counterparts, our inspection tasks were less effective, with marginal improvements over the baselines.

\begin{table}[t]
	\renewcommand{\arraystretch}{1.1}
	\footnotesize
	\centering
	\caption{\textbf{Different scripts:} phishing rates per URL type.}
	\label{tab:performance_language_url}
	\rowcolors{2}{white}{gray!10}
	\begin{tabularx}{\linewidth}{@{}r|XXXXX@{}}
		\toprule
		& \textsc{sub} & \textsc{first} & \textsc{last} & \textsc{path} & \textsc{squat} \\ \midrule

        \textbf{Germany} - Click & 26.3\% & 28.8\% & 24.2\% & 29.8\% & - \\
        Highlight & 25.7\% & - & 38.6\% & 36.2\% & - \\
        Type & 21.2\% & 35.3\% & 20.0\% & 44.4\% & 7.7\% \\ \midrule

        \textbf{Japan} - Click & 48.7\% & 71.4\% & 76.7\% & 80.8\% & - \\
        Highlight & 53.8\% & - & 52.8\% & 51.9\% & - \\
        Type & 57.7\% & 72.5\% & 53.3\% & 52.8\% & 41.3\% \\

		\bottomrule
	\end{tabularx}
\end{table}

\paragraph{Task performance}
We report the performance of the inspection tasks per URL type for the German and Japanese studies in Table~\ref{tab:performance_language_url}.
This data gives us further insights into the poor performance of the Japanese participants: we observe that the clicking task was highly ineffective for impersonation at the beginning and end of the domain.
We reflect that in these cases, the task proposed both the legitimate domain (e.g., \texttt{example.com}) and the phishing one (\texttt{example-login.com})---it is possible that these participants interpreted the task with excessive compliance and selected the (correct) phishing domain instead of reporting it.
This suspicion is further supported by the high phishing rates on the typing task for impersonations at the beginning of the domain---one that requires also pre-existing knowledge of the correct domain to be solved correctly, as discussed in Section~\ref{sec:design.tasks}.
One further observation is the demographic imbalance of the Japanese study, featuring older participants.
\new{While in our US experiment we did not observe correlations between phishing falling and age, older users derived limited benefits from our mechanisms (see Section~\ref{sec:results.demographics}) and are known in the literature to be more susceptible to phishing~\cite{sheng2010falls}.}
Overall, our results suggest that the effectiveness of our mechanism is influenced by familiarity with the script of the URLs; further, they highlight the importance of wording and framing in the design of these types of countermeasures~\cite{franz2021sok,bauer2013warning}.

	\section{Discussion}
	\label{sec:discussion}
	\subsection{Approach Limitations}
\label{sec:discussion.approach}

\paragraph{\new{Knowledge of URLs}}
\new{Inspecting the URL alone cannot help with opaque URLs (e.g., URL shorteners and redirections~\cite{gupta2014bit}) as well as legitimate services hosting malicious content (e.g., a Google Drive document containing malicious links, or a survey service asking for credentials) where the user needs to leverage knowledge and context.
Indeed, nowadays the user needs to know the correct domain for their desired service on the Internet.
For well-known websites and the ones encountered frequently, this is less of a concern than for lesser-known services: URLs pointing to not-so-well-known domains should trigger higher user scrutiny. 
Note that our approach helped heighten user attention also against cloud-based services, URLs more complicated to parse, and out-of-context URLs (all frequently used by phishers; see Section~\ref{sec:results.more_complex_urls}) and prompted users to (re)think.}

\paragraph{Practical Tradeoffs}
An inherent tradeoff between security and usability is in how our approach checks the solution of the tasks.
The ``brand'' of the service might be in a subdomain (\texttt{drive.google.com})---therefore, it makes sense for the user answer to include the subdomain, which should not be considered an error.
However, this might decrease the security of our approach for services that host user-controlled content in subdomains, e.g., consider a phishing URL leading to \texttt{drive.google.com} where the attacker hosts something impersonating \textit{another} service hosted on the same domain such as \texttt{mail.google.com}.
Requiring only the domain as an answer to the task would not protect users as we intend.
We can mitigate the first issue by allowing also subdomain(s) to be part of the answer; the second issue is more challenging to address, as it would require deciding when to require subdomains (e.g., for Google in our example) and when not to (e.g., for a service that does not host user-generated content).

\subsection{Study Limitations}

\paragraph{Generalization}
The demographics of the U.S. and German studies are skewed toward younger, more tech-savvy participants, which might not be representative of the general population.
The study had a short time restriction and was limited to one single session, therefore, how our approach would fare with repeated use must be further investigated.
In our study, we only tested a limited number of handcrafted URLs. 
We did so to ensure quality of the tested URLs, and to reduce one source of variance in our study; further, we tested all the common structures of phishing URLs in their basic structure: longer or more complex URLs were not tested but are expected to fall into one of our tested categories.

\paragraph{Roleplay setting}
Our study incorporated a roleplay setting, which raises several questions regarding participants' (i) motivations, (ii) familiarity with the role, and (iii) realism of the setting.
Regarding motivation, while we did not tie participants' rewards to their performance, some might have felt pressured to perform ``well''---however, the clear goal of the study was fully revealed only after debriefing.
Further, the incentive of participants to complete the study fast is similar to employees in a company who want to manage their email as fast as possible~\cite{greene2018user}.
Finally, we leveraged participants' previous knowledge and thus offered customized and realistic emails and roles based on their experience, and a familiar UI.

\paragraph{Biases}
Participants in different groups might have been differently biased towards the true nature of the experiment, and thus involuntarily nudged towards paying more attention.
Indeed, non-control participants saw countermeasures upon clicking on links and might have understood that the study was about correctly classifying phishing URLs: \new{this might have increased the false positive rates}.
However, this does not impact the comparisons between the baselines and our approach.
Another potential source of bias is that the legitimate URLs employed in the study were overall slightly simpler than the phishing URLs, which might have directed participants towards suspecting phishing when seeing more complicated URLs.
However, this reflects the reality of phishing URLs being more complex than legitimate ones~\cite{althobaiti2021don}.

\paragraph{Data quality}
Finally, we reflect on the quality of the data collected on the online platforms.
We decided not to employ \textit{attention checks} despite their popularity in online studies because they are not recommended in Prolific, do not seem to increase data quality~\cite{tang2022replication}, and for security-related studies they might bias participants into paying more attention than they would in real life~\cite{hauser2015sa}.
Furthermore, we observed good data quality on this platform~\cite{albert2023comparing,douglas2023data,tang2022replication} with very high completion rates, realistic solving times, and low error rates in our study.
Finally, we checked the agreement of participants' gender and age between our questionnaire and the data they provided to Prolific, and only excluded 4 due to mismatches.
		
	\section{Related Work}
	\label{sec:relwork}
	\paragraph{Design of warnings and security UIs}
Warning design is an active area of research, both for physical products~\cite{wogalter2002based} and digital interfaces~\cite{franz2021sok,schaub2015design}.
Design principles for this special type of communication derive from theoretical models of human communication and information processing~\cite{wogalter2018communication}, mental models~\cite{bravo2010bridging}, or from empirical studies of how users interact with warnings, e.g., for SSL warnings in web browsers~\cite{akhawe2013alice,sunshine2009crying,reeder2018experience}, or of privacy notices~\cite{mcdonald2009comparative}.
This research lead to the creation of guidelines for security warnings and UIs~\cite{schaub2015design,franz2021sok,bauer2013warning}.
Effective warnings should be salient~\cite{egelman2008you,reeder2018experience}, concise and accurate~\cite{aneke2019designing,bauer2013warning}, contextual to what triggered them~\cite{petelka2019put}, and attract attention both through design elements~\cite{franz2021sok} and through requiring user interaction to proceed~\cite{felt2015improving,akhawe2013alice,aneke2019designing,bravo2013your}, as users otherwise tend to spend little time on security-relevant indicators~\cite{harrison2016individual,neupane2015multi}.
Habituation and desensitization due to excessive exposure and predictability of the warnings are also a major concern to address~\cite{sunshine2009crying,krol2012don}.

\paragraph{Teaching URLs to users}
Users are generally not very proficient at parsing modern URLs~\cite{albakry2020url,reynolds2020measuring}.
This is especially true for obfuscated and long URLs, where users struggle to understand their structure, and for URLs that impersonate familiar brands by placing their names in some parts of it~\cite{reynolds2020measuring}.
Therefore, several works have proposed tools and games to teach users how URLs are composed and to recognize phishing URLs~\cite{althobaiti2018faheem,canova2015learn,kumaraguru2010teaching}, uncovering features that are most helpful to users~\cite{althobaiti2021don}.
These proposals leverage presentation elements to explain how to divide a URL into its parts~\cite{althobaiti2018faheem,canova2015learn}, as users of all levels of technical proficiency otherwise struggle with reading URLs without help~\cite{albakry2020url}.
They also focus on providing tips and heuristics to recognize phishing URLs~\cite{kumaraguru2010teaching}, and use gamification to make the learning process more engaging~\cite{kumaraguru2010teaching,canova2015learn}.
The main drawback of these support UIs is that they struggle to give users transferable knowledge, as performance can drop after the UI is not available anymore~\cite{althobaiti2018faheem}.

\paragraph{Related UIs}
Domain highlighting is one of the main techniques used to help users understand URLs, by showing the domain part of the URL in a different color or font weight~\cite{volkamer2017user,lin2011does}; however, it is only effective for users with good technical knowledge~\cite{lin2011does}.
Another approach is augmenting existing interfaces to show more indicators, e.g., the sender's name and time of sending~\cite{nicholson2017can}, or the URL's age and popularity~\cite{althobaiti2021don}.
However, all these approaches are passive and thus easy to ignore~\cite{felt2015improving,dhamija2006phishing}; furthermore, increasing the amount of information in passive warnings does not improve phishing detection~\cite{lain2021phishing,zheng2022presenting}.
To help users focus on the URL, studies investigated inhibitive warnings by enforcing delays while a tooltip presents more information about the clicked link~\cite{volkamer2017user}, or requiring to click on it again ~\cite{petelka2019put}, but these can still be prone to habituation as users can click through these warnings without paying attention.
\new{Tasks similar to ours have been successfully explored in the context of untrusted applications~\cite{bravo2013your}, where users were required to retype the name of the application publisher to detect impersonation attacks.
Therefore, it is worth investigating whether these tasks translate to URLs, as they have richer semantics (e.g., components), phishers employ different types of deception, and users have different understanding and mental models.}

	\section{Conclusion}
	\label{sec:conclusion}
	In this paper, we presented \textit{URL inspection tasks}, a novel approach to help users detect phishing URLs in emails.
\new{Our active approach, recommended as a sporadic countermeasure in more sensitive environments such as corporate settings, reduced the victimization rate of participants in the study from 75\% to 25\% and providing strong protection against hard-to-spot typosquatting URLs.}
The effectiveness of our approach comes from a design that follows the guidelines and best practices in warning and security UIs~\cite{schaub2015design,franz2021sok,bauer2013warning} employing contextual, active tasks that help users pay attention, combined with the intention verification aspect.

Our results also offer insights into why users are susceptible to phishing URLs.
The difference in victimization rates between the control group, which includes users with a standard browser-based email client, and the group that interacted with our tasks, highlights the need for better presentation.
This indicates that URLs should be displayed more prominently, as participants often recognized deception while completing the tasks. 
Additionally, we show that up to 50\% of the participants who were initially unsuccessful in solving the tasks were aided by the notification of intention mismatch, demonstrating the need for better education regarding URL structures.
Finally, our design and study highlight that there still exists a gap between technical indicators and users' intentions (e.g., the domain \texttt{example.com} and whether it identifies the intended ``Example'' online service) that needs to be investigated further.
One possible direction is to explore whether our tasks can be simplified and abstracted away from technical indicators and whether this approach might impact security.

Potential future research directions are investigating more user-friendly interventions and better URL education methods; further examining non-native Latin script readers; and adapting our tasks to mobile platforms.

	\section*{Ethics Considerations}
	\label{sec:ethics}
	Our study was approved by the IRB of our institution.
Participants electronically signed a consent form describing the nature of our study and the data we would collect: their answers to the questionnaires, their demographic information provided by the platform, and their interactions with the study platform. All data was stored pseudonymously.
While our initial study description did not explicitly mention participants they would be exposed to phishing, this is a commonly used method in most phishing studies~\cite{resnik2018ethics,thomopoulos2023methodologies} to avoid excessive priming.
The participants were debriefed after completing the study with the full description, and is confirmed to incur only minimal risks~\cite{finn2007designing}, also confirmed by our IRB classifying our study as minimal risk.
Participants were appropriately remunerated for their time with a payment matching the highest minimum wage in their country.

We took further countermeasures to ensure participants' safety: the discomfort of being exposed to phishing emails was mitigated by the roleplay setting and their assigned fictitious identity.
Furthermore, their task was limited to clicking on links---there was no interaction with simulated phishing websites or other potentially harmful content.
Additionally, the phishing URLs we provided did not offer an easy way for participants to actually visit them (as our environment was preventing navigation); however, to protect participants that might transcribe or copy-paste them into their browsers, we constantly monitored all URLs to ensure they were offline during the duration of the study.

    \section*{Open Science}
    The anonymized data recorded from the experiment and the code used to analyze it and generate the figures and tables presented in this paper are available at \url{https://zenodo.org/records/14737023}.
	
	\bibliographystyle{IEEEtran}
	\bibliography{bibliography}

\begin{thebibliography}{10}
\providecommand{\url}[1]{#1}
\csname url@samestyle\endcsname
\providecommand{\newblock}{\relax}
\providecommand{\bibinfo}[2]{#2}
\providecommand{\BIBentrySTDinterwordspacing}{\spaceskip=0pt\relax}
\providecommand{\BIBentryALTinterwordstretchfactor}{4}
\providecommand{\BIBentryALTinterwordspacing}{\spaceskip=\fontdimen2\font plus
\BIBentryALTinterwordstretchfactor\fontdimen3\font minus \fontdimen4\font\relax}
\providecommand{\BIBforeignlanguage}[2]{{%
\expandafter\ifx\csname l@#1\endcsname\relax
\typeout{** WARNING: IEEEtran.bst: No hyphenation pattern has been}%
\typeout{** loaded for the language `#1'. Using the pattern for}%
\typeout{** the default language instead.}%
\else
\language=\csname l@#1\endcsname
\fi
#2}}
\providecommand{\BIBdecl}{\relax}
\BIBdecl

\bibitem{lin2022phish}
X.~Lin, P.~Ilia, S.~Solanki, and J.~Polakis, ``Phish in sheep's clothing: Exploring the authentication pitfalls of browser fingerprinting,'' in \emph{31st USENIX Security Symposium (USENIX Security 22)}, 2022, pp. 1651--1668.

\bibitem{al2022covid}
A.~F. Al-Qahtani and S.~Cresci, ``The covid-19 scamdemic: A survey of phishing attacks and their countermeasures during covid-19,'' \emph{IET Information Security}, vol.~16, no.~5, pp. 324--345, 2022.

\bibitem{mink2022deepphish}
J.~Mink, L.~Luo, N.~M. Barbosa, O.~Figueira, Y.~Wang, and G.~Wang, ``Deepphish: Understanding user trust towards artificially generated profiles in online social networks,'' in \emph{Proc. of USENIX Security}, 2022.

\bibitem{cofense2023}
Cofense, ``Urls 4x more likely than phishing attachments to reach users,'' \url{https://cofense.com/blog/urls-4x-more-likely-than-phishing-attachments-to-reach-users/}, 2023.

\bibitem{aslan2020comprehensive}
{\"O}.~A. Aslan and R.~Samet, ``A comprehensive review on malware detection approaches,'' \emph{IEEE access}, vol.~8, pp. 6249--6271, 2020.

\bibitem{bravo2010bridging}
C.~Bravo-Lillo, L.~F. Cranor, J.~Downs, and S.~Komanduri, ``Bridging the gap in computer security warnings: A mental model approach,'' \emph{IEEE Security \& Privacy}, vol.~9, no.~2, pp. 18--26, 2010.

\bibitem{vishwanath2018suspicion}
A.~Vishwanath, B.~Harrison, and Y.~J. Ng, ``Suspicion, cognition, and automaticity model of phishing susceptibility,'' \emph{Communication research}, vol.~45, no.~8, pp. 1146--1166, 2018.

\bibitem{oest2020phishtime}
A.~Oest, Y.~Safaei, P.~Zhang, B.~Wardman, K.~Tyers, Y.~Shoshitaishvili, and A.~Doup{\'e}, ``$\{$PhishTime$\}$: Continuous longitudinal measurement of the effectiveness of anti-phishing blacklists,'' in \emph{29th USENIX Security Symposium (USENIX Security 20)}, 2020, pp. 379--396.

\bibitem{sheng2009empirical}
S.~Sheng, B.~Wardman, G.~Warner, L.~Cranor, J.~Hong, and C.~Zhang, ``An empirical analysis of phishing blacklists,'' 2009.

\bibitem{greene2018user}
K.~K. Greene, M.~Steves, M.~Theofanos, J.~Kostick \emph{et~al.}, ``User context: an explanatory variable in phishing susceptibility,'' in \emph{in Proc. 2018 Workshop Usable Security}, 2018.

\bibitem{vishwanath2011people}
A.~Vishwanath, T.~Herath, R.~Chen, J.~Wang, and H.~R. Rao, ``Why do people get phished? testing individual differences in phishing vulnerability within an integrated, information processing model,'' \emph{Decision Support Systems}, vol.~51, no.~3, pp. 576--586, 2011.

\bibitem{albakry2020url}
S.~Albakry, K.~Vaniea, and M.~K. Wolters, ``What is this url's destination? empirical evaluation of users' url reading,'' in \emph{Proceedings of the 2020 CHI Conference on Human Factors in Computing Systems}, 2020, pp. 1--12.

\bibitem{reynolds2020measuring}
J.~Reynolds, D.~Kumar, Z.~Ma, R.~Subramanian, M.~Wu, M.~Shelton, J.~Mason, E.~Stark, and M.~Bailey, ``Measuring identity confusion with uniform resource locators,'' in \emph{Proceedings of the 2020 CHI Conference on Human Factors in Computing Systems}, 2020, pp. 1--12.

\bibitem{lin2011does}
E.~Lin, S.~Greenberg, E.~Trotter, D.~Ma, and J.~Aycock, ``Does domain highlighting help people identify phishing sites?'' in \emph{Proceedings of the SIGCHI Conference on Human Factors in Computing Systems}, 2011, pp. 2075--2084.

\bibitem{xiong2017domain}
A.~Xiong, R.~W. Proctor, W.~Yang, and N.~Li, ``Is domain highlighting actually helpful in identifying phishing web pages?'' \emph{Human factors}, vol.~59, no.~4, pp. 640--660, 2017.

\bibitem{dhamija2006phishing}
R.~Dhamija, J.~D. Tygar, and M.~Hearst, ``Why phishing works,'' in \emph{Proceedings of the SIGCHI conference on Human Factors in computing systems}, 2006, pp. 581--590.

\bibitem{lain2021phishing}
D.~Lain, K.~Kostiainen, and S.~Capkun, ``Phishing in organizations: Findings from a large-scale and long-term study,'' in \emph{IEEE S\&P 2022}, 2022.

\bibitem{althobaiti2021don}
K.~Althobaiti, N.~Meng, and K.~Vaniea, ``I don’t need an expert! making url phishing features human comprehensible,'' in \emph{Proceedings of the 2021 CHI Conference on Human Factors in Computing Systems}, 2021, pp. 1--17.

\bibitem{volkamer2017user}
M.~Volkamer, K.~Renaud, B.~Reinheimer, and A.~Kunz, ``User experiences of torpedo: Tooltip-powered phishing email detection,'' \emph{Computers \& Security}, vol.~71, pp. 100--113, 2017.

\bibitem{petelka2019put}
J.~Petelka, Y.~Zou, and F.~Schaub, ``Put your warning where your link is: Improving and evaluating email phishing warnings,'' in \emph{Proceedings of the 2019 CHI conference on human factors in computing systems}, 2019, pp. 1--15.

\bibitem{zeng2021winding}
Y.~Zeng, X.~Chen, T.~Zang, and H.~Tsang, ``Winding path: Characterizing the malicious redirection in squatting domain names,'' in \emph{Passive and Active Measurement: 22nd International Conference, PAM 2021, Virtual Event, March 29--April 1, 2021, Proceedings 22}.\hskip 1em plus 0.5em minus 0.4em\relax Springer, 2021, pp. 93--107.

\bibitem{purkait2014empirical}
S.~Purkait, S.~K. De, and D.~Suar, ``An empirical investigation of the factors that influence internet user’s ability to correctly identify a phishing website,'' \emph{Information Management \& Computer Security}, 2014.

\bibitem{nicholson2017can}
J.~Nicholson, L.~Coventry, and P.~Briggs, ``Can we fight social engineering attacks by social means? assessing social salience as a means to improve phish detection,'' in \emph{Thirteenth Symposium on Usable Privacy and Security (SOUPS 2017)}, 2017, pp. 285--298.

\bibitem{neupane2015multi}
A.~Neupane, M.~L. Rahman, N.~Saxena, and L.~Hirshfield, ``A multi-modal neuro-physiological study of phishing detection and malware warnings,'' in \emph{Proceedings of the 22nd ACM SIGSAC Conference on Computer and Communications Security}, 2015, pp. 479--491.

\bibitem{wang2012research}
J.~Wang, T.~Herath, R.~Chen, A.~Vishwanath, and H.~R. Rao, ``Research article phishing susceptibility: An investigation into the processing of a targeted spear phishing email,'' \emph{IEEE transactions on professional communication}, vol.~55, no.~4, pp. 345--362, 2012.

\bibitem{krol2012don}
K.~Krol, M.~Moroz, and M.~A. Sasse, ``Don't work. can't work? why it's time to rethink security warnings,'' in \emph{2012 7th international conference on risks and security of internet and systems (CRiSIS)}.\hskip 1em plus 0.5em minus 0.4em\relax IEEE, 2012, pp. 1--8.

\bibitem{bravo2013your}
C.~Bravo-Lillo, S.~Komanduri, L.~F. Cranor, R.~W. Reeder, M.~Sleeper, J.~Downs, and S.~Schechter, ``Your attention please: Designing security-decision uis to make genuine risks harder to ignore,'' in \emph{Proceedings of the Ninth Symposium on Usable Privacy and Security}, 2013, pp. 1--12.

\bibitem{li2007usability}
L.~Li and M.~Helenius, ``Usability evaluation of anti-phishing toolbars,'' \emph{Journal in Computer Virology}, vol.~3, pp. 163--184, 2007.

\bibitem{schaub2015design}
F.~Schaub, R.~Balebako, A.~L. Durity, and L.~F. Cranor, ``A design space for effective privacy notices,'' in \emph{Eleventh symposium on usable privacy and security (SOUPS 2015)}, 2015, pp. 1--17.

\bibitem{bauer2013warning}
L.~Bauer, C.~Bravo-Lillo, L.~Cranor, and E.~Fragkaki, ``Warning design guidelines,'' Carnegie Mellon University, Tech. Rep. CMU-CyLab-13-002, 2013.

\bibitem{franz2021sok}
A.~Franz, V.~Zimmermann, G.~Albrecht, K.~Hartwig, C.~Reuter, A.~Benlian, and J.~Vogt, ``$\{$SoK$\}$: Still plenty of phish in the sea—a taxonomy of $\{$User-Oriented$\}$ phishing interventions and avenues for future research,'' in \emph{Seventeenth Symposium on Usable Privacy and Security (SOUPS 2021)}, 2021, pp. 339--358.

\bibitem{searles2023empirical}
A.~Searles, Y.~Nakatsuka, E.~Ozturk, A.~Paverd, G.~Tsudik, and A.~Enkoji, ``An empirical study \& evaluation of modern $\{$CAPTCHAs$\}$,'' in \emph{32nd usenix security symposium (usenix security 23)}, 2023, pp. 3081--3097.

\bibitem{tupsamudre2019everything}
H.~Tupsamudre, A.~K. Singh, and S.~Lodha, ``Everything is in the name--a url based approach for phishing detection,'' in \emph{International symposium on cyber security cryptography and machine learning}.\hskip 1em plus 0.5em minus 0.4em\relax Springer, 2019, pp. 231--248.

\bibitem{aung2019survey}
E.~S. Aung, C.~T. Zan, and H.~Yamana, ``A survey of url-based phishing detection,'' in \emph{DEIM forum}, 2019, pp. G2--3.

\bibitem{canova2015nophish}
G.~Canova, M.~Volkamer, C.~Bergmann, and B.~Reinheimer, ``Nophish app evaluation: lab and retention study,'' in \emph{NDSS workshop on usable security}, 2015.

\bibitem{tang2022replication}
J.~Tang, E.~Birrell, and A.~Lerner, ``Replication: How well do my results generalize now? the external validity of online privacy and security surveys,'' in \emph{Eighteenth symposium on usable privacy and security (SOUPS 2022)}, 2022, pp. 367--385.

\bibitem{douglas2023data}
B.~D. Douglas, P.~J. Ewell, and M.~Brauer, ``Data quality in online human-subjects research: Comparisons between mturk, prolific, cloudresearch, qualtrics, and sona,'' \emph{Plos one}, vol.~18, no.~3, p. e0279720, 2023.

\bibitem{hauser2015sa}
D.~J. Hauser and N.~Schwarz, ``It’sa trap! instructional manipulation checks prompt systematic thinking on “tricky” tasks,'' \emph{Sage Open}, vol.~5, no.~2, p. 2158244015584617, 2015.

\bibitem{sheng2010falls}
S.~Sheng, M.~Holbrook, P.~Kumaraguru, L.~F. Cranor, and J.~Downs, ``Who falls for phish? a demographic analysis of phishing susceptibility and effectiveness of interventions,'' in \emph{Proceedings of the SIGCHI conference on human factors in computing systems}, 2010, pp. 373--382.

\bibitem{gupta2014bit}
N.~Gupta, A.~Aggarwal, and P.~Kumaraguru, ``bit. ly/malicious: Deep dive into short url based e-crime detection,'' in \emph{2014 APWG Symposium on Electronic Crime Research (eCrime)}.\hskip 1em plus 0.5em minus 0.4em\relax IEEE, 2014, pp. 14--24.

\bibitem{albert2023comparing}
D.~A. Albert and D.~Smilek, ``Comparing attentional disengagement between prolific and mturk samples,'' \emph{Scientific Reports}, vol.~13, no.~1, p. 20574, 2023.

\bibitem{wogalter2002based}
M.~S. Wogalter, V.~C. Conzola, and T.~L. Smith-Jackson, ``Research-based guidelines for warning design and evaluation,'' \emph{Applied ergonomics}, vol.~33, no.~3, pp. 219--230, 2002.

\bibitem{wogalter2018communication}
M.~S. Wogalter, ``Communication-human information processing (c-hip) model,'' in \emph{Forensic human factors and ergonomics}.\hskip 1em plus 0.5em minus 0.4em\relax CRC Press, 2018, pp. 33--49.

\bibitem{akhawe2013alice}
D.~Akhawe and A.~P. Felt, ``Alice in warningland: a $\{$Large-Scale$\}$ field study of browser security warning effectiveness,'' in \emph{22nd USENIX Security Symposium (USENIX Security 13)}, 2013, pp. 257--272.

\bibitem{sunshine2009crying}
J.~Sunshine, S.~Egelman, H.~Almuhimedi, N.~Atri, and L.~F. Cranor, ``Crying wolf: An empirical study of ssl warning effectiveness.'' in \emph{USENIX security symposium}.\hskip 1em plus 0.5em minus 0.4em\relax Montreal, Canada, 2009, pp. 399--416.

\bibitem{reeder2018experience}
R.~W. Reeder, A.~P. Felt, S.~Consolvo, N.~Malkin, C.~Thompson, and S.~Egelman, ``An experience sampling study of user reactions to browser warnings in the field,'' in \emph{Proceedings of the 2018 CHI conference on human factors in computing systems}, 2018, pp. 1--13.

\bibitem{mcdonald2009comparative}
A.~M. McDonald, R.~W. Reeder, P.~G. Kelley, and L.~F. Cranor, ``A comparative study of online privacy policies and formats,'' in \emph{International Symposium on Privacy Enhancing Technologies Symposium}.\hskip 1em plus 0.5em minus 0.4em\relax Springer, 2009, pp. 37--55.

\bibitem{egelman2008you}
S.~Egelman, L.~F. Cranor, and J.~Hong, ``You've been warned: an empirical study of the effectiveness of web browser phishing warnings,'' in \emph{Proceedings of the SIGCHI Conference on Human Factors in Computing Systems}, 2008, pp. 1065--1074.

\bibitem{aneke2019designing}
J.~Aneke, C.~Ardito, and G.~Desolda, ``Designing an intelligent user interface for preventing phishing attacks,'' in \emph{IFIP Conference on Human-Computer Interaction}.\hskip 1em plus 0.5em minus 0.4em\relax Springer, 2019, pp. 97--106.

\bibitem{felt2015improving}
A.~P. Felt, A.~Ainslie, R.~W. Reeder, S.~Consolvo, S.~Thyagaraja, A.~Bettes, H.~Harris, and J.~Grimes, ``Improving ssl warnings: Comprehension and adherence,'' in \emph{Proceedings of the 33rd annual ACM conference on human factors in computing systems}, 2015, pp. 2893--2902.

\bibitem{harrison2016individual}
B.~Harrison, E.~Svetieva, and A.~Vishwanath, ``Individual processing of phishing emails: How attention and elaboration protect against phishing,'' \emph{Online Information Review}, 2016.

\bibitem{althobaiti2018faheem}
K.~Althobaiti, K.~Vaniea, and S.~Zheng, ``Faheem: Explaining urls to people using a slack bot,'' in \emph{Symposium on digital behaviour intervention for cyber security}, 2018, pp. 1--8.

\bibitem{canova2015learn}
G.~Canova, M.~Volkamer, C.~Bergmann, R.~Borza, B.~Reinheimer, S.~Stockhardt, and R.~Tenberg, ``Learn to spot phishing urls with the android nophish app,'' in \emph{IFIP World Conference on Information Security Education}.\hskip 1em plus 0.5em minus 0.4em\relax Springer, 2015, pp. 87--100.

\bibitem{kumaraguru2010teaching}
P.~Kumaraguru, S.~Sheng, A.~Acquisti, L.~F. Cranor, and J.~Hong, ``Teaching johnny not to fall for phish,'' \emph{ACM Transactions on Internet Technology (TOIT)}, vol.~10, no.~2, pp. 1--31, 2010.

\bibitem{zheng2022presenting}
S.~Zheng and I.~Becker, ``Presenting suspicious details in user-facing e-mail headers does not improve phishing detection,'' in \emph{USENIX Symposium on Usable Privacy and Security (SOUPS)}.\hskip 1em plus 0.5em minus 0.4em\relax USENIX Association, 2022.

\bibitem{resnik2018ethics}
D.~B. Resnik and P.~R. Finn, ``Ethics and phishing experiments,'' \emph{Science and engineering ethics}, vol.~24, no.~4, pp. 1241--1252, 2018.

\bibitem{thomopoulos2023methodologies}
G.~Thomopoulos, D.~Lyras, and C.~Fidas, ``Methodologies and ethical considerations in phishing research: A comprehensive review,'' in \emph{Proceedings of the 2nd International Conference of the ACM Greek SIGCHI Chapter}, 2023, pp. 1--10.

\bibitem{finn2007designing}
P.~Finn and M.~Jakobsson, ``Designing ethical phishing experiments,'' \emph{IEEE Technology and Society Magazine}, vol.~26, no.~1, pp. 46--58, 2007.

\end{thebibliography}

	\appendix
	\section{Study Materials}
\label{sec:appendix}

\subsection{Baseline Tasks}
\label{sec:appendix.baselines}
\begin{figure}[t]
	\centering
	\begin{subfigure}[b]{0.85\linewidth}
		\centering
		\includegraphics[width=0.5\linewidth]{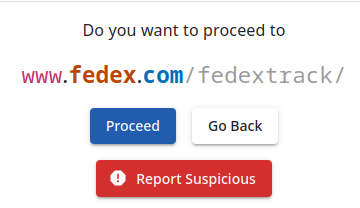}
		\caption{Passive Task.}
		\label{fig:baselines.passive}
	\end{subfigure}\\
	\begin{subfigure}[b]{\linewidth}
		\centering
		\includegraphics[width=0.9\linewidth]{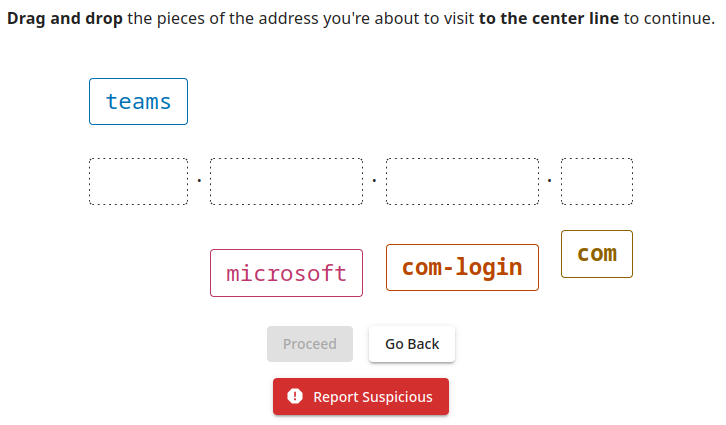}
		\caption{Active Task.}
		\label{fig:baselines.active}
	\end{subfigure}
	\caption{Baseline tasks used in the study.}
	\label{fig:baselines}
\end{figure}

We show in Figure~\ref{fig:baselines} the two baseline tasks we compared against in this study: Figure~\ref{fig:baselines.passive} shows the passive task, where participants were simply asked to review the URL and confirm it.
Figure~\ref{fig:baselines.active} shows the active task, where participants were asked to drag the URL components to the center line and then confirm whether they wanted to visit the page.

\subsection{URLs Used in the Study}
\label{sec:appendix.urls}
We report all the legitimate and phishing URLs for each service in Table~\ref{tab:all_urls}.
For each service, we also show the path fragment we used in the URLs, which was the same for both types of URLs. To ensure a high degree of realism, the path fragments were chosen from common ones that the legitimate services use.

\begin{table*}[t]
	\renewcommand{\arraystretch}{1.8}
    \scriptsize
	\centering
	\caption{All the legitimate and phishing URLs used in the study. Phishing domains also featured the same path fragment of the legitimate URL. \textsc{path} URLs had the legitimate one as first part of their path.}
	\label{tab:all_urls}
	\rowcolors{2}{white}{gray!10}
	\begin{tabularx}{\linewidth}{@{}X|XXXXX@{}}
		\toprule
		\textbf{Service Name} & \textsc{sub} & \textsc{first} & \textsc{last} & \textsc{path} & \textsc{squat} \\ \midrule

		{\footnotesize\bf Sharepoint} \vspace{0.5em} \newline futuracom-my.sharepoint.com \newline /personal/taylor\_futuracom\_ & {sharepoint.com-login.com} & {futuracom.sharepoint-login.com} & {futuracom.login-my-sharepoint.com} & {futuracom.secure-login.com} & {futuracom-my.sharep0int.com} \\{\footnotesize\bf Google Drive} \vspace{0.5em} \newline drive.google.com \newline /drive/folders/1t8FLJdJzDSOsMFYv & {drive.google.com-login.com} & {drive.google-login.com} & {drive.login-google.com} & {secure-login.com} & {drive.googie.com} \\{\footnotesize\bf Microsoft Teams} \vspace{0.5em} \newline teams.microsoft.com \newline /\_\#/conversations/?ctx=chat & {teams.microsoft.com-login.com} & {teams.microsoft-login.com} & {teams.login-microsoft.com} & {secure-login.com} & {teams.mircosoft.com} \\{\footnotesize\bf Facebook} \vspace{0.5em} \newline www.facebook.com \newline /login/?next= & {www.facebook.com-login.com} & {www.facebook-login.com} & {www.profile.login-facebook.com} & {www.secure-login.com} & {www.facebok.com} \\{\footnotesize\bf LinkedIn} \vspace{0.5em} \newline www.linkedin.com \newline /in/futuracom/recent-activity/all & {www.linkedin.com-login.com} & {www.linkedin-login.com} & {www.profile.login-linkedin.com} & {www.secure-login.com} & {www.linkedln.com} \\{\footnotesize\bf PayPal} \vspace{0.5em} \newline www.paypal.com \newline /myaccount/activities/details/8BC09211LP421880G & {www.paypal.com-login.com} & {www.paypal-login.com} & {www.login-paypal.com} & {www.secure-login.com} & {www.paypai.com} \\{\footnotesize\bf FedEx} \vspace{0.5em} \newline www.fedex.com \newline /fedextrack/?trknbr=400394482105 & {www.fedex.com-login.com} & {www.fedex-login.com} & {www.login-fedex.com} & {www.secure-tracking.com} & {www.fed-ex.com} \\
		\midrule
		\multicolumn{6}{l}{\textbf{\new{Follow-up study URLs}}} \\
		\midrule
		{\footnotesize\bf \new{Sharepoint (Hard)}} \vspace{0.5em} \newline {futuracom.cloudapp.{\textcolor{white}-}azure.com} \newline /personal/taylor\_futuracom\_ & {futuracom.cloudapp.azure.{\textcolor{white}-}com-login.com} & {futuracom.cloudapp.azure-login.com} & {https://futuracom.login-cloudapp-azure.com} & {futuracom.secure-login.com} & {https://futuracom-cloudapp.4zure.com} \\
		{\footnotesize\bf \new{Google Drive (Hard)}} \vspace{0.5em} \newline {futuracom.spreadsheets0.{\textcolor{white}-}google.com} \newline /file/1t8FLJdJzDSOsMFYv & {futuracom.spreadsheets0.{\textcolor{white}-}google.com-login.com} & {futuracom.spreadsheets0.{\textcolor{white}-}google-login.com} & {futuracom.spreadsheets0.{\textcolor{white}-}login-google.com} & {futuracom.secure-login.com} & futuracom.spreadsheets0.{\textcolor{white}-}googie.com \\
	{\footnotesize\bf \new{Invoicing system}} \vspace{0.5em} \newline {admin.internal.futuracom.org} \newline /invoice/314766 & {admin.internal.futuracom.org-login.org} & {admin.internal.futuracom-login.org} & {admin.internal.login-futuracom.org} & {admin.internal.secure-login.org} & - \\
	{\footnotesize\bf \new{Intranet}} \vspace{0.5em} \newline {intranet.futuracom.org} \newline /docs/2025/internal-restructuring & {intranet.futuracom.org-login.org} & {intranet.futuracom-login.org} & {intranet.login-futuracom.org} & {intranet.secure-login.org} & - \\
	\multicolumn{6}{l}{{\footnotesize\bf Unknown URL} (both for legitimate and phishing emails)} \\
	\multicolumn{6}{l}{192.175.32.86.bc.googleusercontent.com/doc/1t8FLJdJzDSOsMFYv} \\
		\bottomrule
    \end{tabularx}
\end{table*}

\begin{figure}[t]
    \centering
    \includegraphics[width=.9\columnwidth]{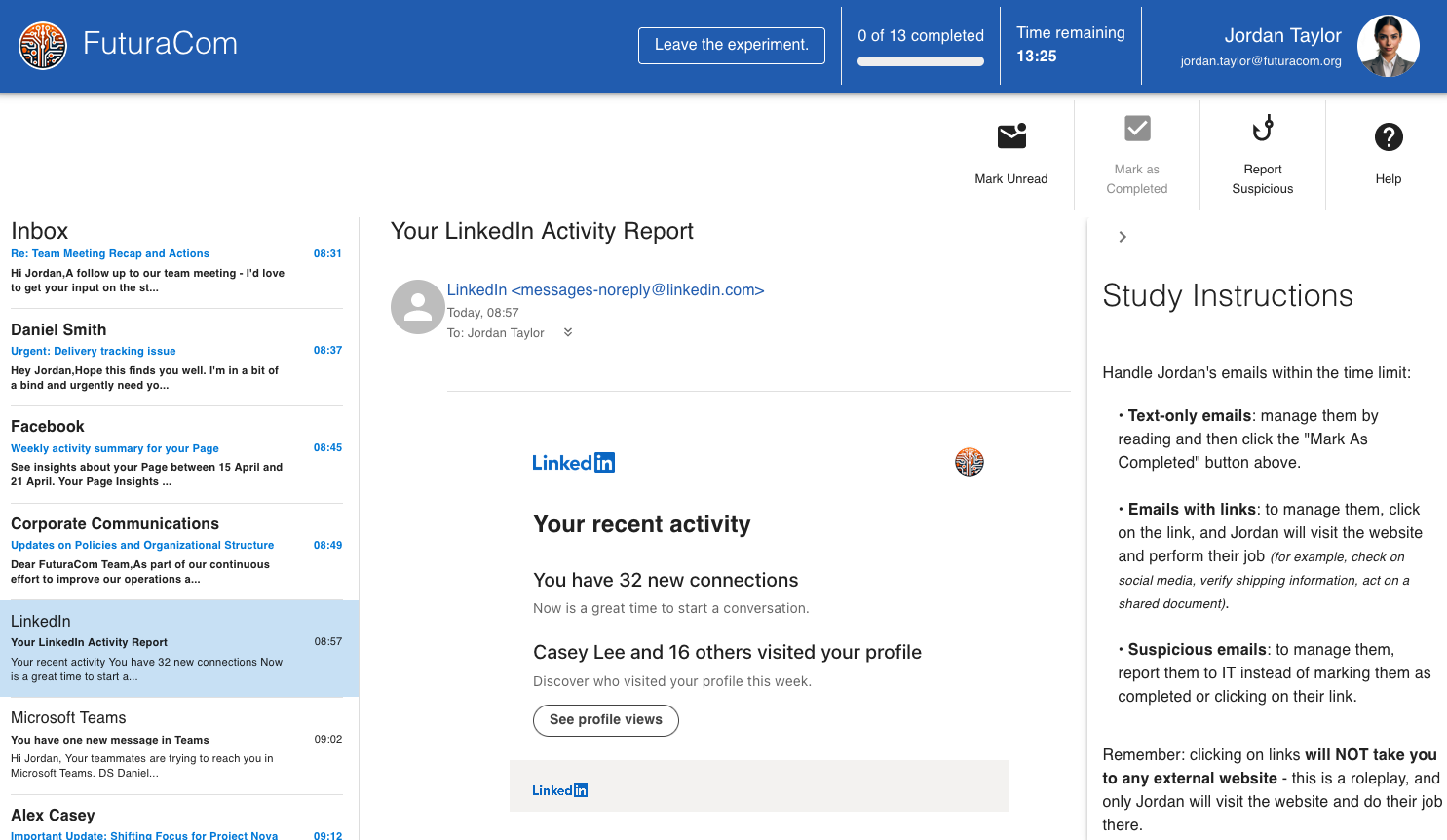}
    \caption{\textbf{The email client participants used in the study}, mimicking the popular Outlook Web App.}
    \label{fig:platform.client}
\end{figure}

\subsection{Experimental Platform}

We show in Figure~\ref{fig:platform.client} the interface of the email client we developed for the study: it features a familiar look-and-feel as well as reminding participants of the study protocol and instructions.

\subsection{Questionnaires}
\label{sec:appendix.post_questionnaire}
Our post-study questionnaire included the following questions:

\noindent \textbf{Common questions to all participants:}
	\begin{compactitem}
		\item \textit{I felt confident in detecting the scam emails by reading.}
		\item \textit{I felt the scam emails were difficult to detect.}
		\item \textit{I felt the legitimate URLs were easy to recognize as such.}
		\item \textit{I felt the scam URLs were difficult to spot from the email.}
	\end{compactitem}
\noindent \textbf{Baseline tasks questions:}
	\begin{compactitem}
		\item \textit{[Re-reading / Having to reorder] the clicked URLs on the confirmation page was useful.}
		\item \textit{I ignored the URL on the page.}
		\item \textit{[Seeing / Reordering] the URLs on the confirmation page helped me decide.}
	\end{compactitem}
\noindent \textbf{Inspection tasks questions:}
	\begin{compactitem}
		\item \textit{The link challenges helped me spot phishing URLs.}
		\item \textit{I did not need the challenges to understand which URLs were phishing.}
		\item \textit{The tool clearly highlighted mistakes I made in reading the URLs.}
		\item \textit{The challenges were in the way of doing my job.}
		\item \textit{The challenge tutorial was clear.}
		\item \textit{The challenge tutorial presented all the information I needed.}
		\item \textit{Coloring the different URL components was useful.}
		\item \textit{I wish the URL was made simpler to read.}
		\item \textit{The challenge presentation was confusing.}
		\item \textit{The challenge to [click / highlight / type] was useful.}
		\item \textit{The challenge to [click / highlight / type] was annoying.}
		\item \textit{The challenge to [click / highlight / type] was difficult.}
	\end{compactitem}

\section{Additional Results}
\label{sec:appendix_results}

\subsection{Accuracy per Demographics}
\label{sec:appendix.demographics_accuracy}

\begin{figure*}[t]
\centering
\begin{subfigure}{0.19\textwidth}
\centering
\includegraphics[width=\textwidth]{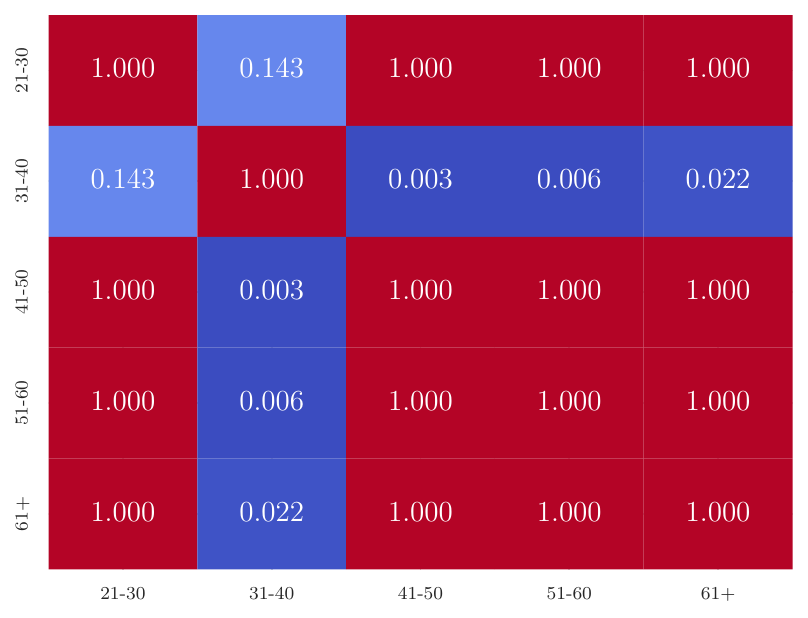}
\caption{Phishing - Passive\\Age}
\label{fig:phishing_baseline_Age}
\end{subfigure}
\begin{subfigure}{0.19\textwidth}
\centering
\includegraphics[width=\textwidth]{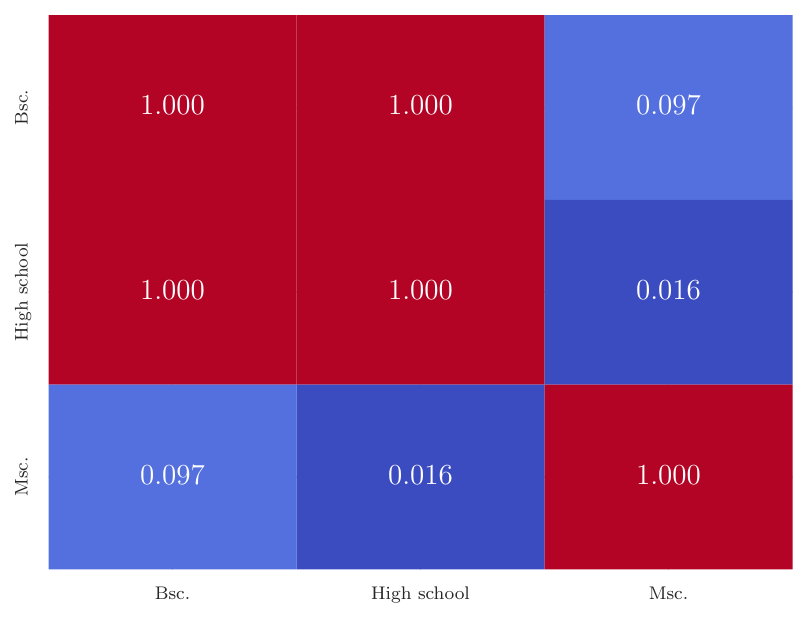}
\caption{Phishing - Passive\\Education}
\label{fig:phishing_baseline_Education}
\end{subfigure}
\begin{subfigure}{0.19\textwidth}
\centering
\includegraphics[width=\textwidth]{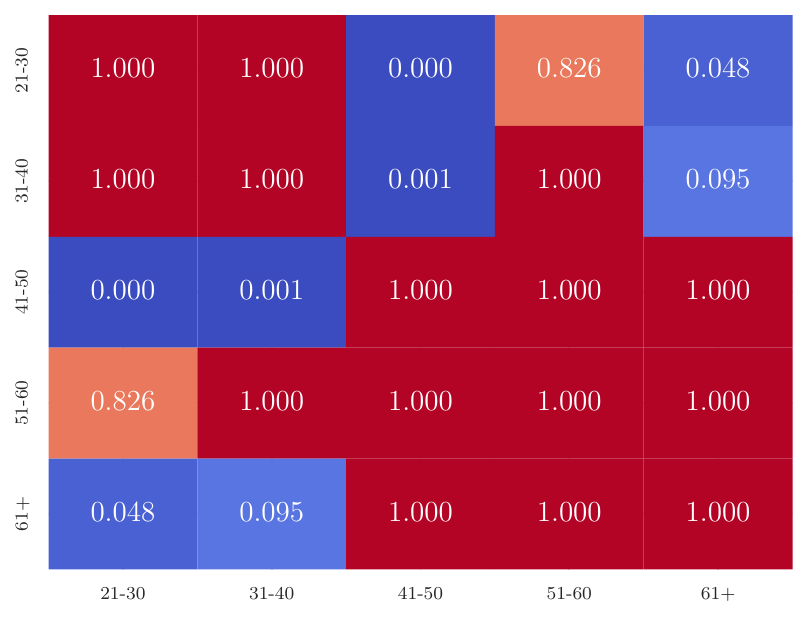}
\caption{Legitimate - Inspection Tasks -- Age}
\label{fig:legitimate_treatment_Age}
\end{subfigure}
\begin{subfigure}{0.19\textwidth}
\centering
\includegraphics[width=\textwidth]{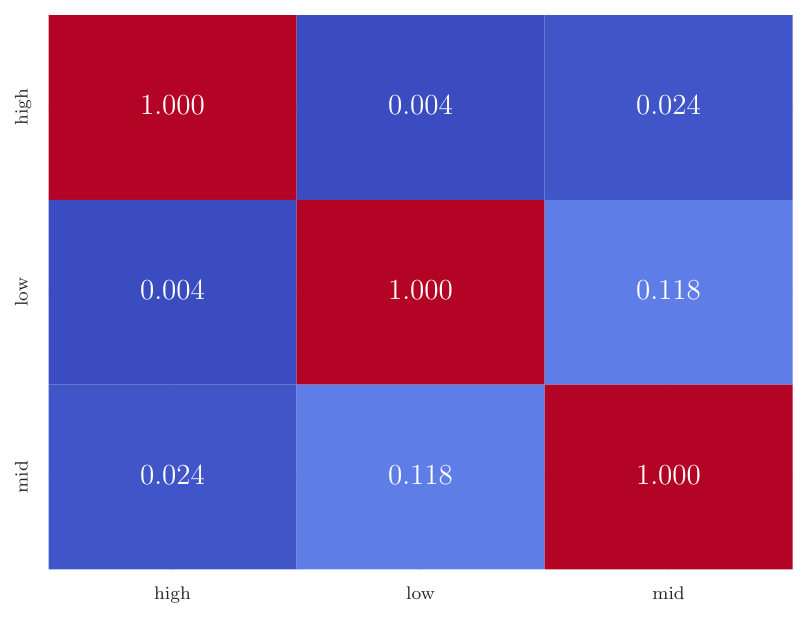}
\caption{Legitimate - Inspection Tasks -- Personal tech use}
\label{fig:legitimate_treatment_Personal tech}
\end{subfigure}
\begin{subfigure}{0.19\textwidth}
\centering
\includegraphics[width=\textwidth]{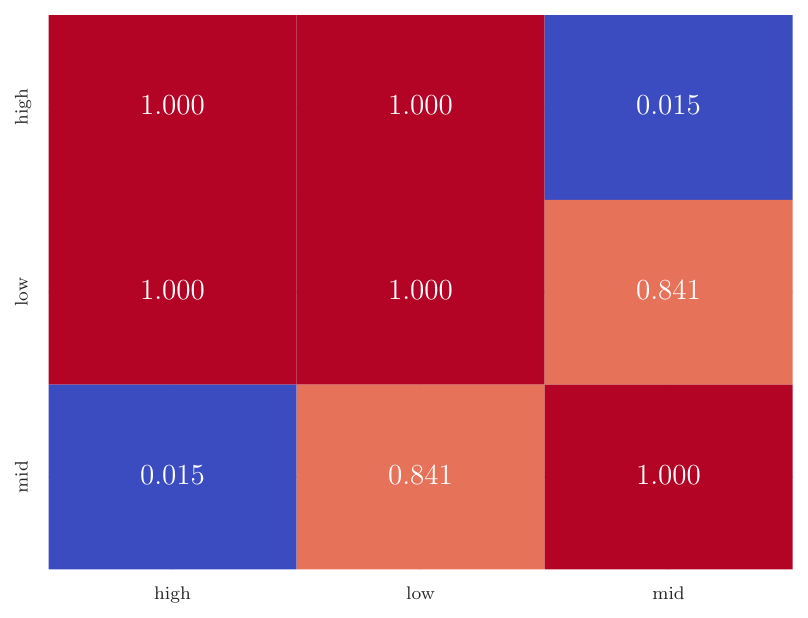}
\caption{Phishing - Inspection Tasks -- Personal tech use}
\label{fig:phishing_treatment_Personal tech}
\end{subfigure}
\caption{Statistical significance of accuracy in legitimate and phishing emails, per mechanism and demographic.}
\label{fig:demographics_accuracy}
\end{figure*}

We report in Figure~\ref{fig:demographics_accuracy} the statistically significant correlations we observed between demographics and task accuracy.

\subsection{Solving Time per Demographics}
\label{sec:appendix.demographics_time}

\begin{figure*}[t]
\centering
\begin{subfigure}{0.19\textwidth}
\centering
\includegraphics[width=\textwidth]{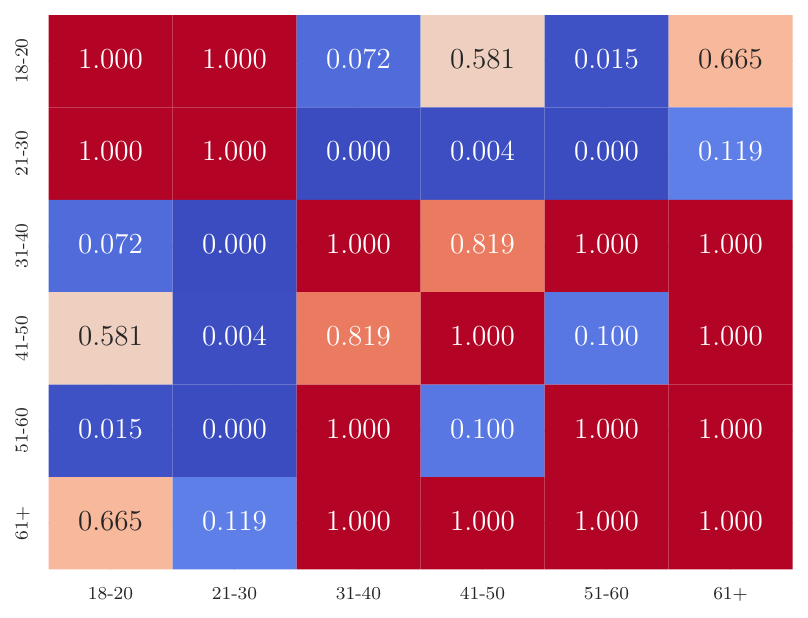}
\end{subfigure}
\begin{subfigure}{0.19\textwidth}
\centering
\includegraphics[width=\textwidth]{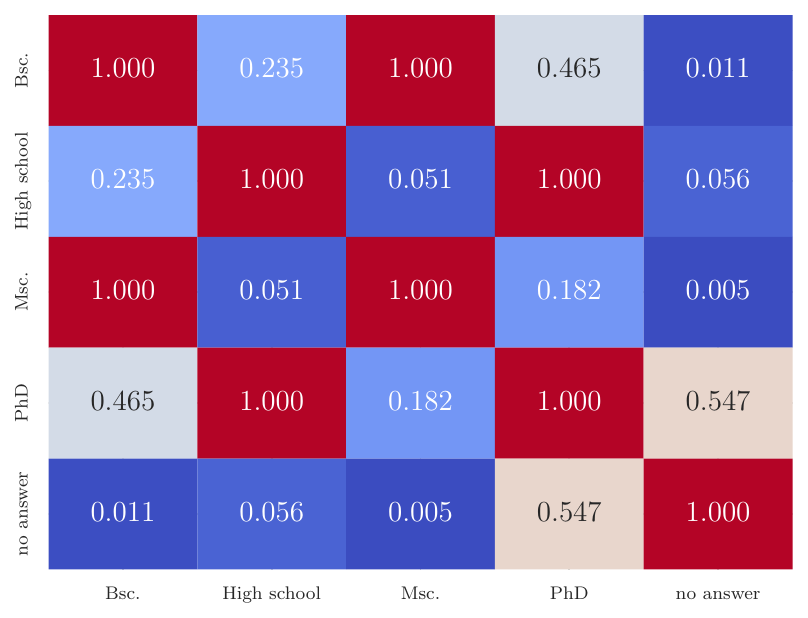}
\end{subfigure}
\begin{subfigure}{0.19\textwidth}
\centering
\includegraphics[width=\textwidth]{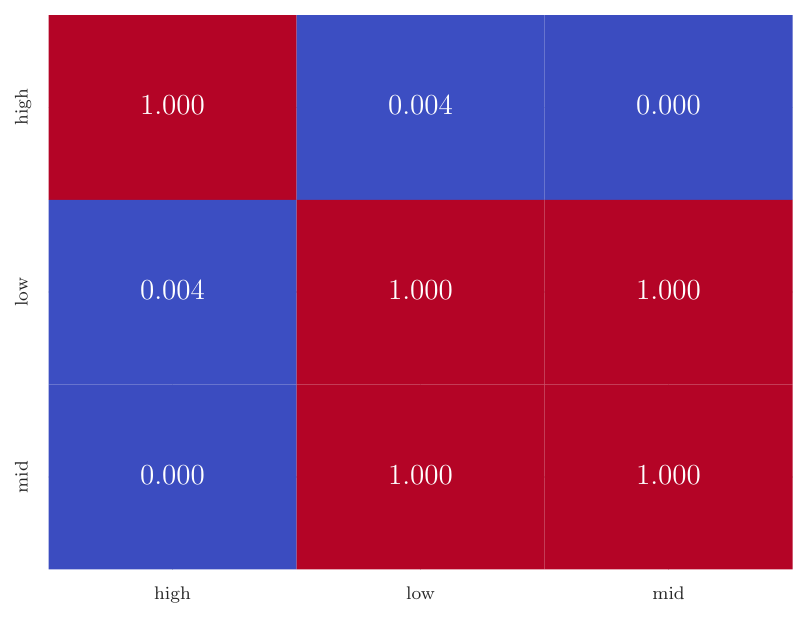}
\end{subfigure}
\begin{subfigure}{0.19\textwidth}
\centering
\includegraphics[width=\textwidth]{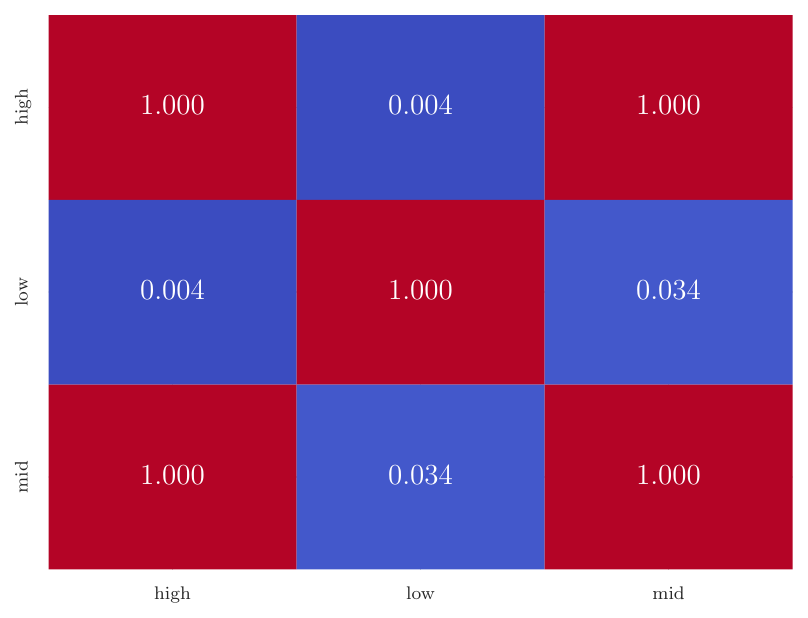}
\end{subfigure}
\begin{subfigure}{0.19\textwidth}
\centering
\includegraphics[width=\textwidth]{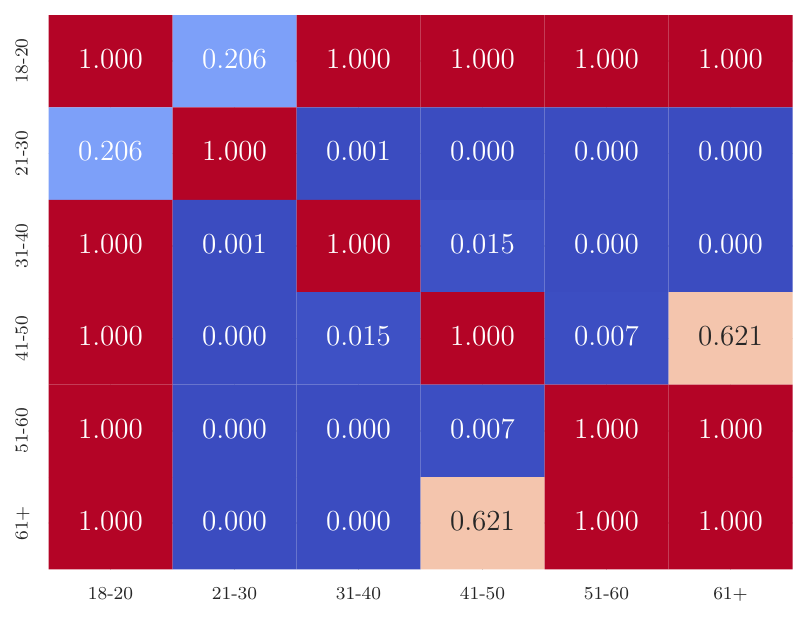}
\end{subfigure}

\begin{subfigure}{0.19\textwidth}
\centering
\includegraphics[width=\textwidth]{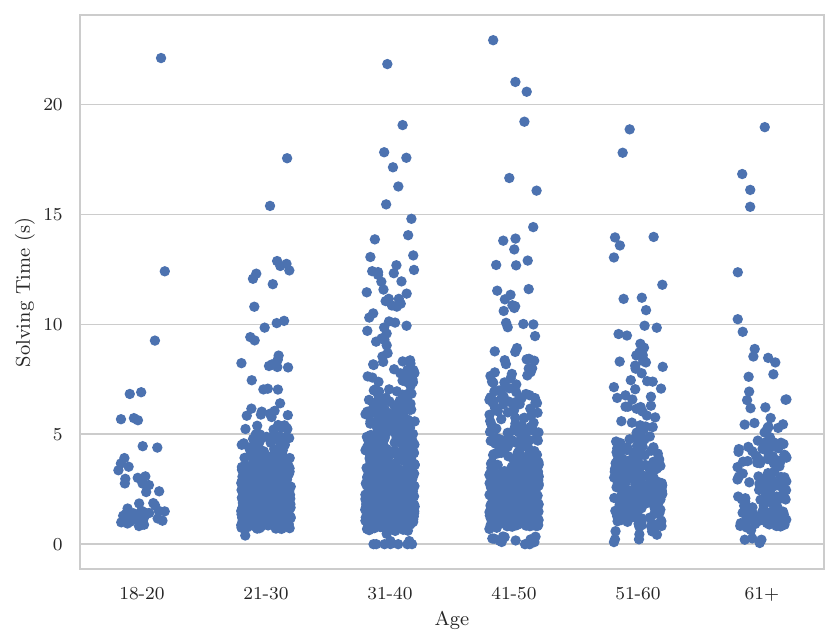}
\caption{Passive - Age\\~}
\label{fig:time_strip_Baseline_Age}
\end{subfigure}
\begin{subfigure}{0.19\textwidth}
\centering
\includegraphics[width=\textwidth]{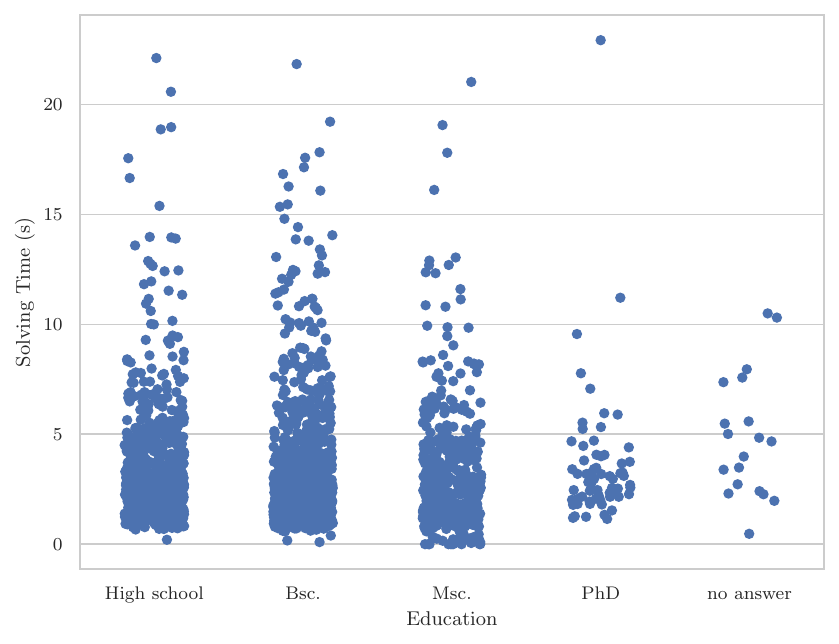}
\caption{Passive - Education\\~}
\label{fig:time_strip_Baseline_Education}
\end{subfigure}
\begin{subfigure}{0.19\textwidth}
\centering
\includegraphics[width=\textwidth]{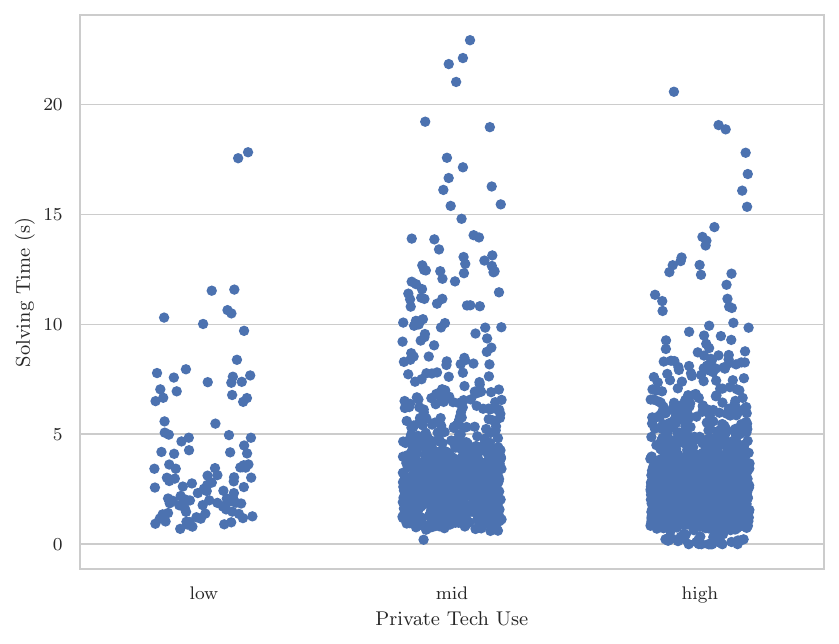}
\caption{Passive - Personal\\Tech Use}
\label{fig:time_strip_Baseline_Private Tech Use}
\end{subfigure}
\begin{subfigure}{0.19\textwidth}
\centering
\includegraphics[width=\textwidth]{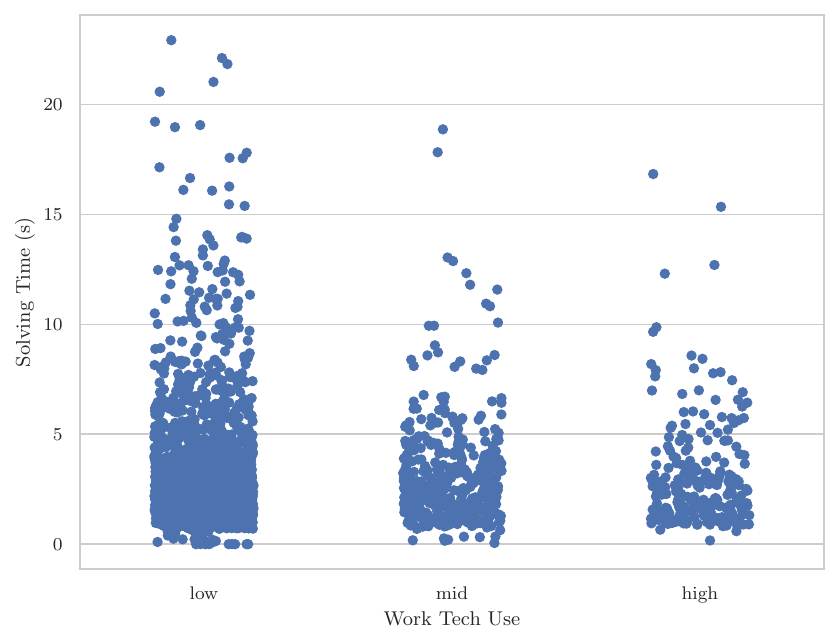}
\caption{Passive - Work\\Tech Use}
\label{fig:time_strip_Baseline_Work Tech Use}
\end{subfigure}
\begin{subfigure}{0.19\textwidth}
\centering
\includegraphics[width=\textwidth]{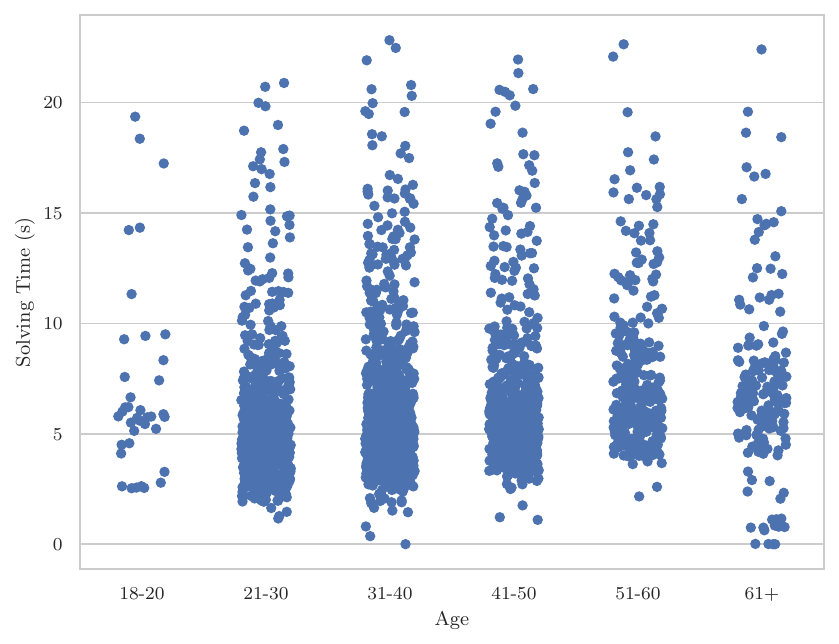}
\caption{Active - Age\\~}
\label{fig:time_strip_Reorder_Age}
\end{subfigure}

\begin{subfigure}{0.19\textwidth}
\centering
\includegraphics[width=\textwidth]{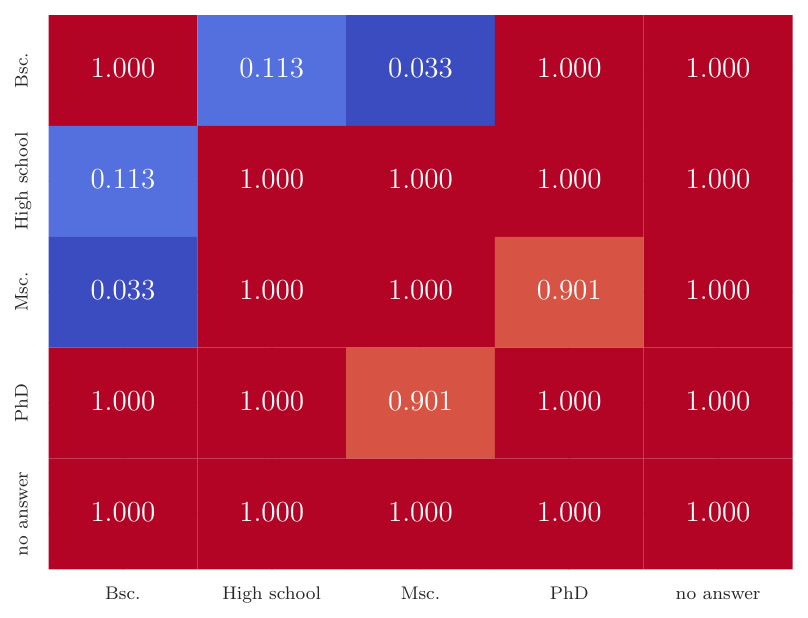}
\end{subfigure}
\begin{subfigure}{0.19\textwidth}
\centering
\includegraphics[width=\textwidth]{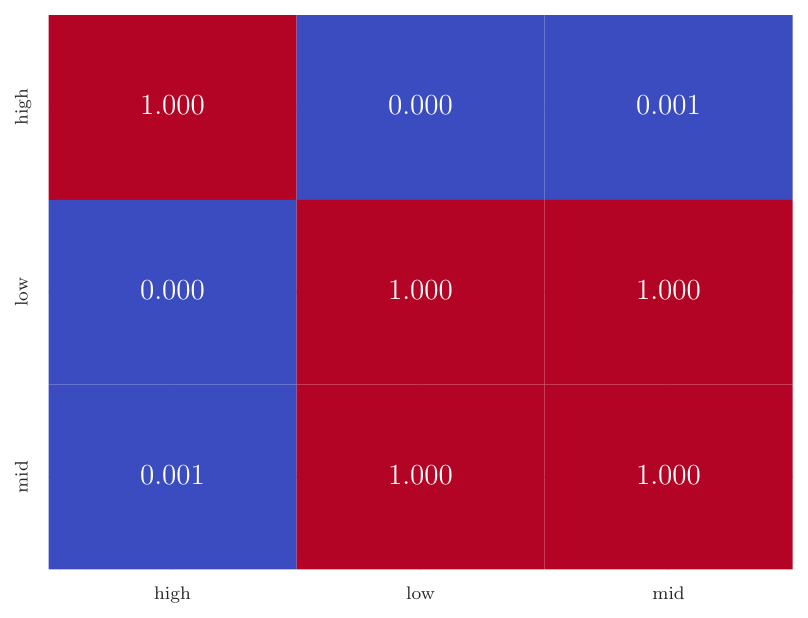}
\end{subfigure}
\begin{subfigure}{0.19\textwidth}
\centering
\includegraphics[width=\textwidth]{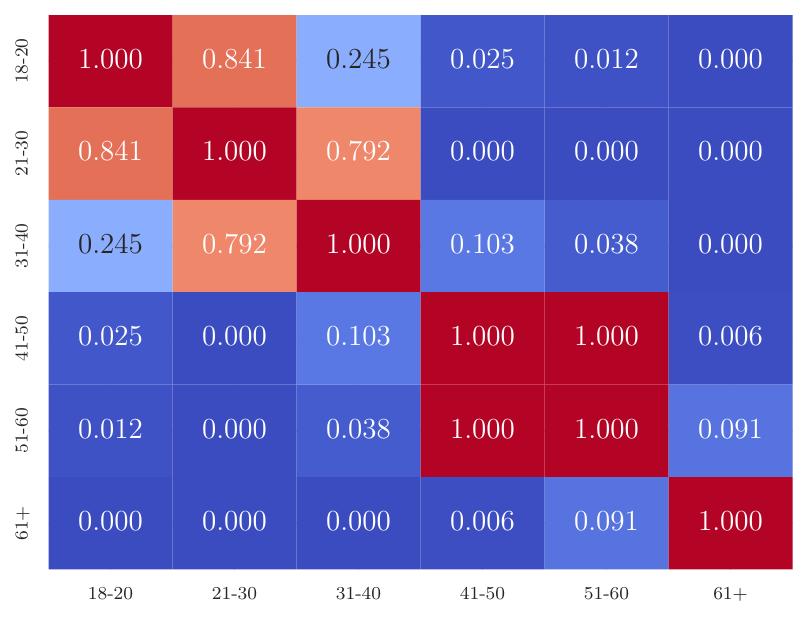}
\end{subfigure}
\begin{subfigure}{0.19\textwidth}
\centering
\includegraphics[width=\textwidth]{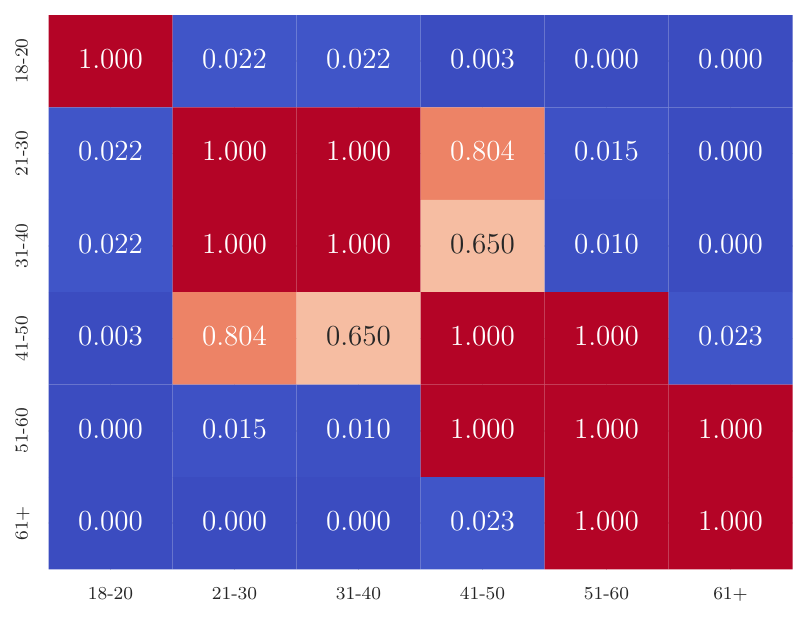}
\end{subfigure}
\begin{subfigure}{0.19\textwidth}
\centering
\includegraphics[width=\textwidth]{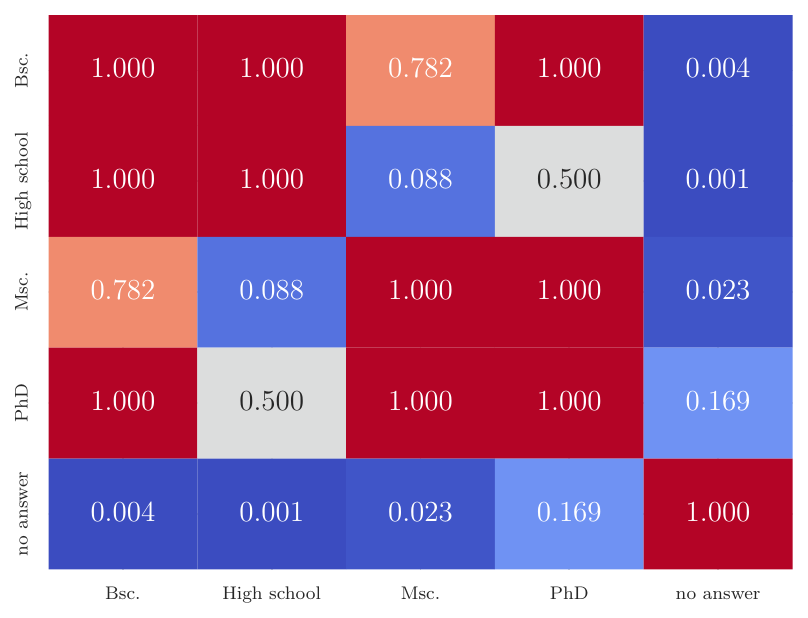}
\end{subfigure}

\begin{subfigure}{0.19\textwidth}
\centering
\includegraphics[width=\textwidth]{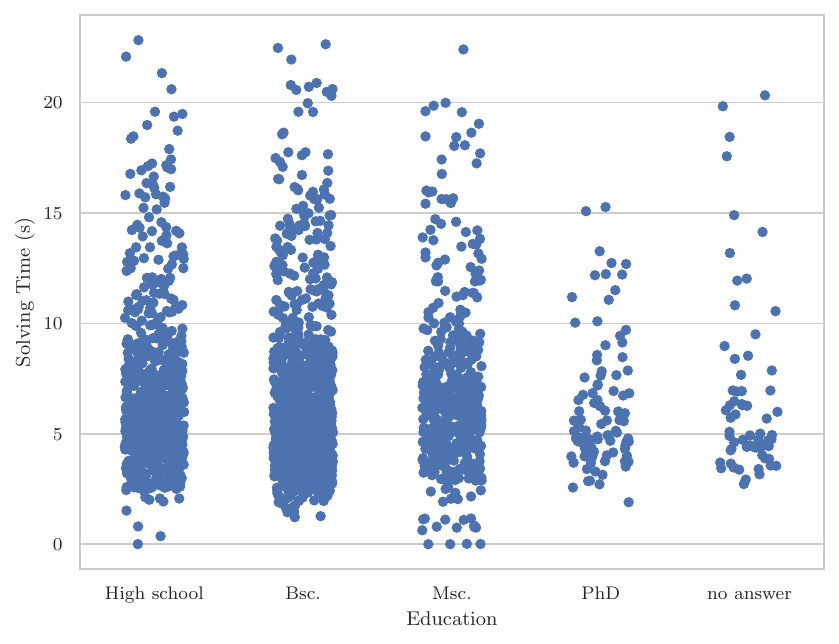}
\caption{Active - Education\\~}
\label{fig:time_strip_Reorder_Education}
\end{subfigure}
\begin{subfigure}{0.19\textwidth}
\centering
\includegraphics[width=\textwidth]{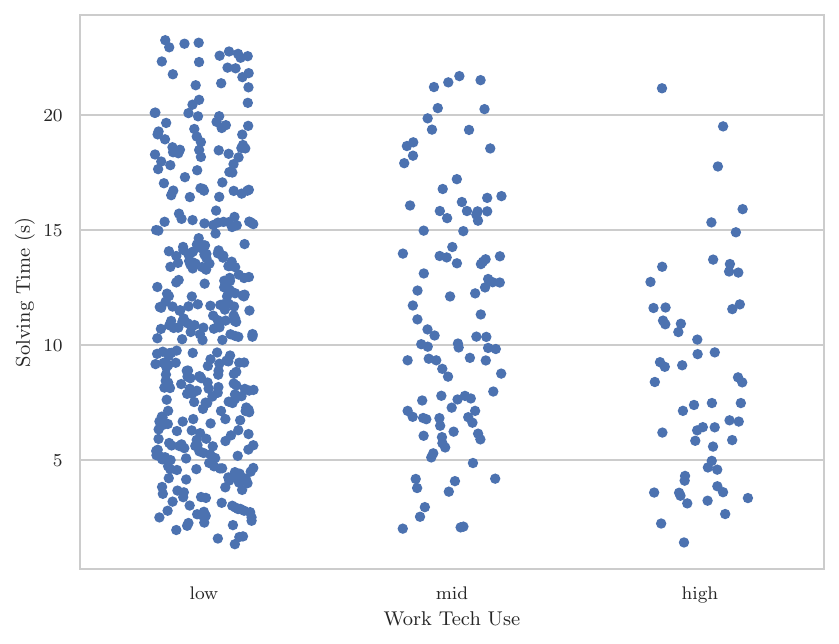}
\caption{Click - Work Tech Use\\~}
\label{fig:time_strip_Click_Work Tech Use}
\end{subfigure}
\begin{subfigure}{0.19\textwidth}
\centering
\includegraphics[width=\textwidth]{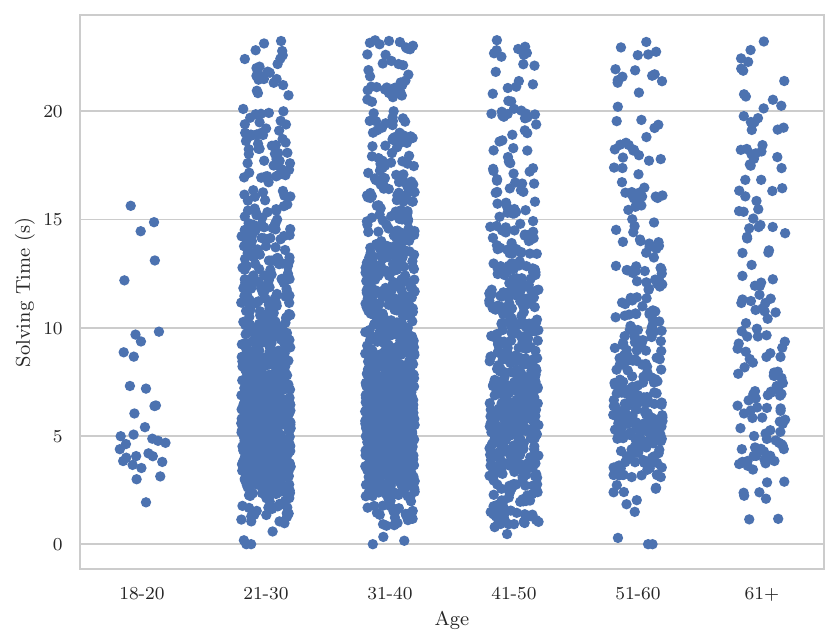}
\caption{Highlight - Age\\~}
\label{fig:time_strip_Highlight_Age}
\end{subfigure}
\begin{subfigure}{0.19\textwidth}
\centering
\includegraphics[width=\textwidth]{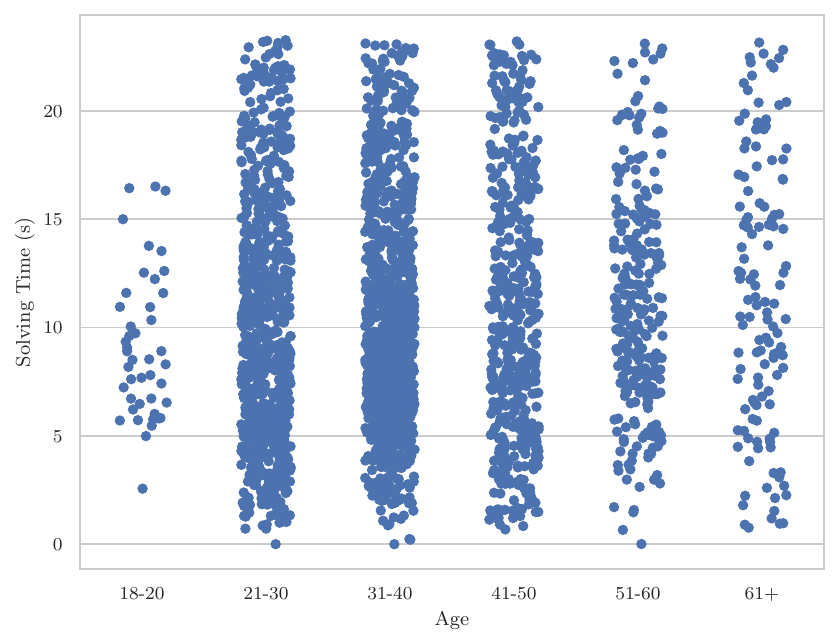}
\caption{Type - Age\\~}
\label{fig:time_strip_Type_Age}
\end{subfigure}
\begin{subfigure}{0.19\textwidth}
\centering
\includegraphics[width=\textwidth]{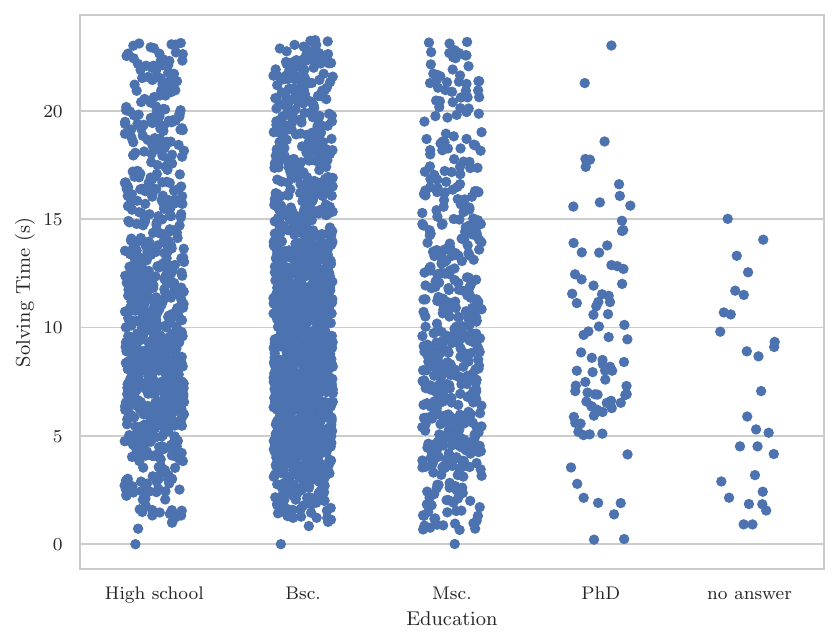}
\caption{Type - Education\\~}
\label{fig:time_Type_Education}
\end{subfigure}
\caption{Statistical significance of solving time per mechanism and demographic.}
\label{fig:demographics_time}
\end{figure*}

We show in Figure~\ref{fig:demographics_time} both a distribution of the solving time per task type and per demographic group for each variable that had a statistically significant correlation, and the correlations between such variables.

\end{document}